\documentclass[aps,prx,twocolumn,superscriptaddress,floatfix]{revtex4-2}

\usepackage{amsmath}
\usepackage{amssymb}
\usepackage{graphicx}
\usepackage{hyperref}
\usepackage{xcolor}
\usepackage{enumitem} 

\begin{document}

\title{Photon echo from lensing of fractional excitations in Tomonaga-Luttinger spin liquid}
\date{\today}

\author{Zi-Long Li}
\affiliation{Institute of Physics, Chinese Academy of Sciences, Beijing 100190, China}
\affiliation{University of Chinese Academy of Sciences, Beijing 100049, China}

\author{Masaki Oshikawa}
\affiliation{Institute for Solid State Physics, University of Tokyo, Kashiwa, Chiba 277-8581, Japan}
\affiliation{Kavli Institute for the Physics and Mathematics of the Universe (WPI), University of Tokyo, Kashiwa, Chiba 277-8583, Japan}

\author{Yuan Wan}
\email{yuan.wan@iphy.ac.cn}
\affiliation{Institute of Physics, Chinese Academy of Sciences, Beijing 100190, China}
\affiliation{University of Chinese Academy of Sciences, Beijing 100049, China}
\affiliation{Songshan Lake Materials Laboratory, Dongguan, Guangdong 523808, China}

\begin{abstract}
We study theoretically the nonlinear optical response of Tomonaga-Luttinger spin liquid in the context of terahertz (THz) two-dimensional coherent spectroscopy (2DCS). Using the gapless phase of the XXZ-type spin chain as an example, we show that its third-order nonlinear magnetic susceptibilities $\chi^{(3)}_{+--+}$ and  $\chi^{(3)}_{-++-}$ exhibit photon echo, where $\pm$ refers to the left/right-hand circular polarization with respect to the $S^z$ axis. The photon echo arises from a ``lensing'' phenomenon in which the wave packets of fractional excitations move apart and then come back toward each other, amounting to a refocusing of the excitations' world lines. Renormalization group irrelevant corrections to the fixed point Hamiltonian result in dispersion and/or damping of the wave packets, which can be sensitively detected by lensing and consequently the photon echo. Our results thus unveil the strength of THz-2DCS in probing the dynamical properties of the collective excitations in a prototypical gapless many-body system.
\end{abstract}

\maketitle

\section{Introduction \label{sec:intro}}

Progress in condensed matter physics is intimately connected to the development of new spectroscopic techniques. Among the many emerging spectroscopies, \emph{ two-dimensional coherent spectroscopy} (2DCS) stands out as a promising tool for investigating strongly correlated systems. The 2DCS use multiple coherent electromagnetic waves to probe the nonlinear optical properties of a sample, thereby producing a two-dimensional spectrum that visualizes the sample's nonlinear response as a function of the frequencies of probing electromagnetic waves~\cite{Mukamel1995,Hamm2011}.

Comparing to the more familiar one-dimensional spectroscopy that probes linear optical properties, the 2DCS reveals not only the optical excitations in a sample but also their relationship. In the infrared and higher frequency range, its ability to diagnose the interplay between optical excitations has been widely leveraged by chemists to unravel the structure of complex molecules and map out the kinetic pathways of chemical reactions~\cite{Mukamel1995,Hamm2011,Cundiff2013}. The advent of terahertz (THz) 2DCS now puts this technique in the right energy window to study many-body phenomena. The THz 2DCS has offered new experimental insights into quantum wells~\cite{Woerner2013}, antiferromagnets~\cite{Lu2017}, and electronic glasses~\cite{Mahmood2020}. On the theory front, it has recently been suggested that the THz 2DCS can resolve the spectral continua formed by optical excitations in several clean and disordered many-body system and characterize their dynamical properties, which would be challenging to accomplish with linear spectroscopy, if at all~\cite{Wan2019,Choi2020,Parameswaran2020,Nandkishore2020,Nguyen2020}.

A main strength of the 2DCS lies in the \emph{photon echo}~\cite{Kurnit1964} signal from the third-order nonlinear optical response $\chi^{(3)}$. The photon echo is an optical analogue of the spin echo in nuclear magnetic resonance (NMR)~\cite{Hahn1950}. Schematically, one may observe the photon echo by exciting the system with three successive short optical pulses and detecting the resulted $\chi^{(3)}$ response (Fig.~\ref{fig:schematic}(a)). Let $\tau$ be the time delay between the second and first pulses, $t_\mathrm{w}$ (the waiting time) be the delay between the third and second, and $t$ be the delay between the time of detection and the last pulse. The photon echo signal appears as a surge of the nonlinear response at $t \approx \tau$ in close analogy with the nuclear magnetic resonance (NMR) spin echo (Fig.~\ref{fig:schematic}(b)).

Similar to its NMR cousin, the photon echo signal can diagnose dissipation in a few-body system by effecting a time-reversal operation --- the system's evolution during the time delay $t$ reverses the evolution occurred during the time delay $\tau$~\cite{Mukamel1995,Hamm2011}. Had the dynamics been unitary, the quantum mechanical phase accumulated in $\tau$ would be completely removed when $t = \tau$, resulting in a perfect rephasing. This perfect rephasing process would produce a photon echo signal at $t=\tau$ regardless the value of $\tau$. Deviation from the perfect rephasing is thus a direct measure of dissipation in the few-body system. Specifically, the decay of echo signal with increasing $\tau$ is a manifestation of the decoherence time ($T_2$ time), whereas the decay of the signal as a function of $t_\mathrm{w}$ probes the population time ($T_1$ time).

This unique ability of diagnosing dissipation makes one naturally wonder if the photon echo could also find its success in strongly correlated many-body systems. Although the latest theoretical inquiries have suggested interesting applications of photon echo in gapped systems~\cite{Wan2019,Choi2020,Nandkishore2020} and disordered systems~\cite{Parameswaran2020}, less attention is paid to clean, gapless many-body systems. Adapting this technique to a gapless many-body setting poses new theoretical challenges. The standard framework for analyzing the photon echo uses the language of energy levels~\cite{Mukamel1995,Hamm2011}, which is best suited for few-body systems with discrete energy spectra. While it is possible to analyze the zero temperature optical response of a gapped system by truncating the Fock space to a subspace containing finite number of excitations~\cite{Babujian2016} and thereby making use of the established framework, such truncation is not permitted in general for gapless systems. Important questions such as the existence of photon echo in gapless strongly-correlated systems, its underlying mechanism, and its features require theoretical investigation.

In this work, we address these questions by studying the nonlinear optical response of a prototypical gapless strongly-correlated system, namely the Tomonaga-Luttinger spin liquid (henceforth ``Luttinger spin liquid" for short)~\cite{Affleck1988,Giamarchi2003}. For concreteness, we consider the Luttinger spin liquid hosted by the XXZ spin chain, which possesses a global $U(1)$ spin rotational symmetry with respect to the spin $z$ axis. We consider exclusively the nonlinear magnetic response $\perp z$ and decompose the electromagnetic wave polarization in the left-handed ($+$) and right-handed ($-$) basis. 

Using the bosonization Hamiltonian at the renormalization group (RG) fixed point, we find that, among the six symmetry-allowed third-order magnetic susceptibilities, $\chi^{(3)}_{+--+}$ and its complex conjugate $\chi^{(3)}_{-++-}$ show photon echo, which appears as a peak on the $t$ axis at $t \approx \tau$. The echo signal possesses a universal asymptotic form, which we obtain analytically and verify numerically. Crucially, the photon echo is ``perfect" in the sense that the signal is a function of $t-\tau$ rather than $t$ and $\tau$ both, resembling the perfectly rephasing photon echo in a few-body system. This implies that the signal measured at a given value of $t-\tau$ is independent of $\tau$. Moreover, the echo signal depends weakly on the waiting time $t_\mathrm{w}$, and saturates when $t_\mathrm{w}\to \infty$. 

This perfect photon echo, although resembles the one due to the prefect rephasing process in few-body systems, comes as a pleasant surprise --- The rephasing process is understood as a result of the optical transitions between discrete energy levels. Here, the rephasing picture does not directly apply as the present system has a continuous energy spectrum. Instead, we trace its origin back to a unique ``lensing" phenomenon of fractional excitations in Luttinger spin liquids (Fig.~\ref{fig:lensing}c): The first THz pulse creates two wave packets of fractional excitations with opposite chirality~\cite{Pham2000}. The second pulse converts the left-moving wave packet into a right-moving one, whereas the third pulse converts the right-moving wave packet into a left-moving one. These two wave packets then meet each other later at $t=\tau$, thereby producing an echo. For the fixed-point Hamiltonian, the phonon modes are exact eigenstates of the Hamiltonian, and their dispersion relation is linear. Consequently, the wave packets of fractional excitations can propagate through the system indefinitely without decay or dispersion. This naturally explains the perfect photon echo, namely the echo signal does not decay as either $\tau$ or $t_\mathrm{w}$ increases.

Our lensing picture immediately suggests that the photon echo is a sensitive diagnostic to the RG-irrelevant perturbations to the fixed point Hamiltonian. These RG-irrelevant corrections give only minor corrections to the most physical quantities at low temperature, and consequently they are often elusive to experimental probes. In the 2DCS, these corrections manifest themselves in the decay of the photon echo thanks to lensing. 

In the XXZ spin chain, the RG-irrelevant corrections include the umklapp terms and higher order gradient terms~\cite{Affleck1988,Giamarchi2003}. The umklapp terms give rise to the dissipation of phonon modes and therefore the decay of the wave packets at finite temperature. As a result, the lensing becomes unattainable when the pulse delay $\tau$ or $t_\mathrm{w}$ exceeds the lifetime of the wave packets. This is manifest as the decay of the echo signal as a function of $\tau$ or $t_\mathrm{w}$, which is analogous to the dissipation-induced decay of the phonon echo in few-body systems mentioned above.

The higher order gradient terms, on the other hand, may result in the decay of the photon echo through a different mechanism. By adding these terms, one may introduce a small curvature to the dispersion relation of the phonon modes while keeping them as the exact eigenstates. The curvature results in the dispersion of the wave packets. We expect the lensing to be ineffective beyond a time scale $\tau_\mathrm{disp}$, at which point the width of the wave packet is comparable with the correlation length. 

This dispersion-induced decay of the photon echo is distinct from the dissipation-induced decay and finds no immediate analogue in few-body systems. We study this decay mechanism on a toy model, namely the harmonic chain, which is a lattice discretization of the fixed point Hamiltonian. The photon echo, when measured at $t=\tau$, decays as a stretched exponential $\sim \exp(-C (\tau/\tau_\mathrm{disp})^{1/2})$, where $C$ is a numerical constant. Meanwhile, the echo signal shows weak dependence on the waiting time $t_\mathrm{w}$ and saturates when $t_\mathrm{w}\to\infty$. We attribute the lack of $t_\mathrm{w}$ dependence to the absence of thermalization in the toy model --- The decay of the photon echo as a function of $t_\mathrm{w}$ reflects the population time. Since the population of the phonon modes cannot relax, its population time is effectively infinity.

To summarize, our analysis shows that the $\chi^{(3)}$ photon echo from the Luttinger spin liquid is a sensitive diagnostic of the RG-irrelevant perturbations to the fixed point Hamiltonian, which are difficult to detect with linear optical spectroscopy. It also uncovers a dispersion-induced photon echo decay mechanism unique to many-body systems. Conceptually, the lensing of fractional excitations is a convenient picture for understanding the photon echo in the Luttinger spin liquid. The lensing picture extends the phase interference picture, commonly invoked for the photon echo in few-body systems~\cite{Mukamel1995,Hamm2011}, from the time domain to the spacetime domain. 

The rest of this work is organized as follows: In Sec.~\ref{sec:setup}, we describe the problem setup. We present the bosonization analysis in Sec..~\ref{sec:bosonization} and the lensing picture in Sec.~\ref{sec:lensing}.  We investigate the dispersion-induced photon echo decay in Sec.~\ref{sec:dispersion}. In Sec.~\ref{sec:discussion}, we point out a few interesting open problems.

\section{Setup \label{sec:setup}}

\begin{figure}
\includegraphics[width = \columnwidth]{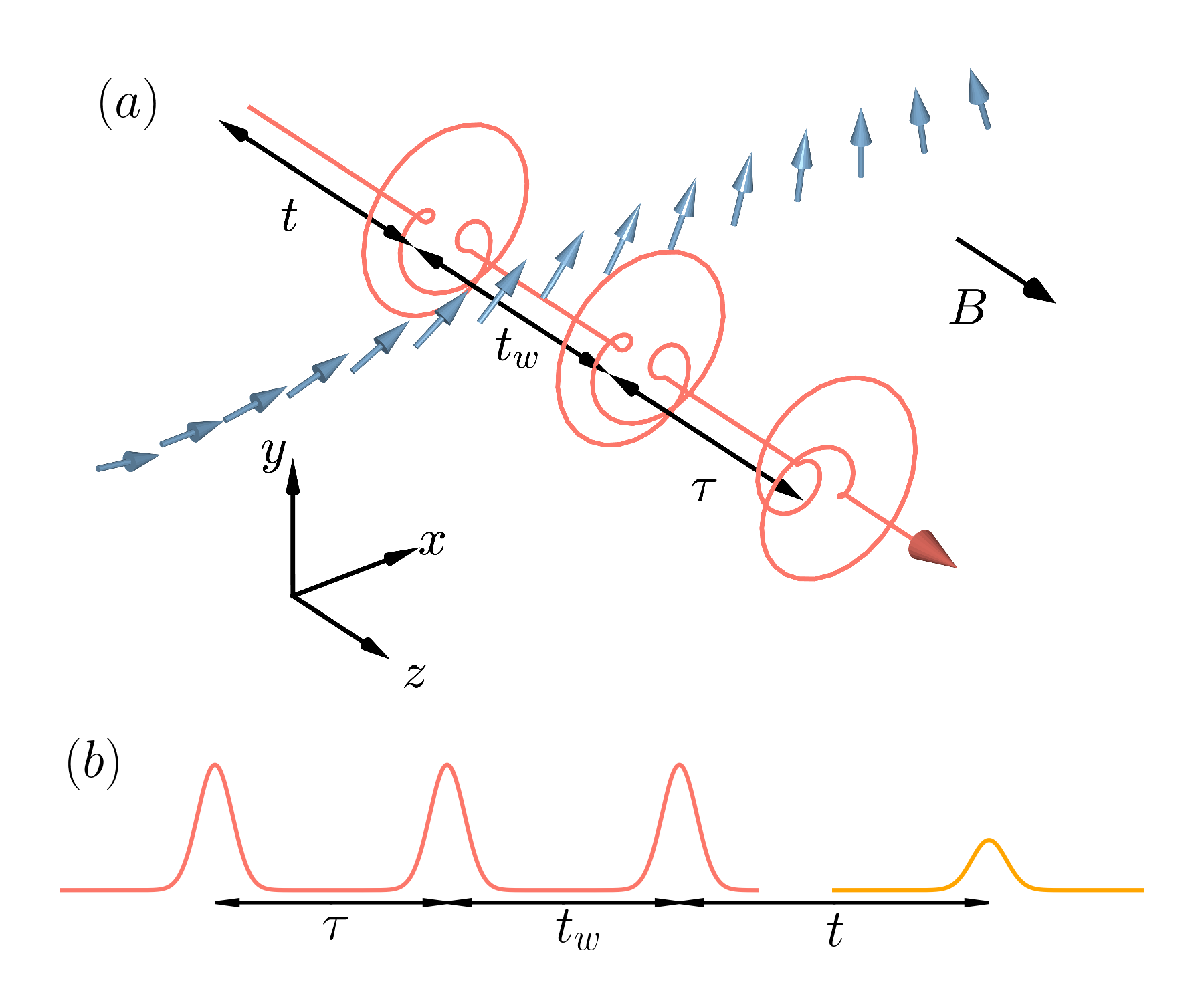}
\caption{(a) The Faraday configuration considered in this work. A magnetic field $B$ is applied in the $z$ axis. Three short electromagnetic pulses with circular polarizations pass through the $S=1/2$ spin chain. The propagation direction is parallel with the spin $z$ axis. The first pulse is right-handed, whereas the second and the third are left-handed. We denote the time delay between the first and the second pulse by $\tau$, the second and the third by $t_\mathrm{w}$, and the third pulse and the time of detection by $t$. (b) The photon echo appears as the surge of nonlinear optical response at the time $t\approx\tau$.}
\label{fig:schematic}
\end{figure}

We consider the $S=1/2$ XXZ spin chain:
\begin{align}
H = \sum_{j} \frac{J_\perp}{2}(S^+_j S^-_{j+1} + h.c.)+ J_z S^z_j S^z_{j+1}-BS^z_j.
\label{eq:xxz_hamil}
\end{align}
$j$ labels the lattice sites. $S^{\pm}_j,S^z$ are the $S=1/2$ spin operators. $J_\perp$ is the exchange constant in the spin $xy$ plane. We shall consider both ``ferromagnetic" ($J_\perp<0$) and ``antiferromagnetic" ($J_\perp>0$) chains. $J_z$ is the exchange constant in the spin $z$ axis. We include the Zeeman term due to an external field $B \parallel z$ (Fig.~\ref{fig:schematic}). Throughout this work, we use the natural units with $\hbar = k_B = \mu = 1$ where $\mu$ is the magnetic moment carried by the spin. We may extend  Eq.~\eqref{eq:xxz_hamil}  by including additional terms so long as they preserve the symmetries. Our analysis is applicable to the Luttinger spin liquid phase of Eq.~\eqref{eq:xxz_hamil} and its extensions.

The 2DCS measures a sample's nonlinear optical response~\cite{Mukamel1995,Hamm2011}.  The electromagnetic wave interacts with an insulating spin system such as Eq.~\eqref{eq:xxz_hamil} primarily by the Zeeman coupling. Therefore, the nonlinear optical response of Eq.~\eqref{eq:xxz_hamil} is chiefly due to its leading-order nonlinear magnetic susceptibility. Note the photon echo that arises from this nonlinear magnetic response closely resembles the NMR spin echo in that both are induced by a sequence of magnetic field pulses, and the former may be justifiably called spin echo as well~\cite{Lu2017}. However, different from the canonical NMR spin echo set up~\cite{Hahn1950}, 2DCS does not require precise control of field pulse area. Here, we adhere to the term ``photon echo" to emphasize this difference from the NMR spin echo.

The Hamiltonian Eq.~\eqref{eq:xxz_hamil} possesses a $U(1)$ spin rotational symmetry with respect to the $z$ axis. Since the total magnetization in $z$ commutes with $H$ and does not evolve in the Heisenberg picture, it is natural to consider the magnetic response in the $xy$ plane. Experimentally, this corresponds to the \emph{Faraday geometry} where the propagation direction of the electromagnetic wave is parallel/anti-parallel to the external field (Fig.~\ref{fig:schematic}a)~\cite{Zvezdin1997}. 

The $U(1)$ symmetry of the Hamiltonian Eq.~\eqref{eq:xxz_hamil} forbids second-order in-plane nonlinear magnetic susceptibilities. The same symmetry allows six third-order in-plane nonlinear magnetic susceptibilities, out of which three are independent: $\chi^{(3)}_{+-+-}$, $\chi^{(3)}_{+--+}$, and $\chi^{(3)}_{++--}$, where $+(-)$ refers to the left (right)-handed circular polarization of the electromagnetic wave. The other three susceptibilities, namely $\chi^{(3)}_{-+-+}$, $\chi^{(3)}_{-++-}$, and $\chi^{(3)}_{--++}$, are related to the former three by complex conjugation and thus offer no new information. Throughout this work, we focus on $\chi^{(3)}_{+--+}$, which exhibits the photon echo, and defer the discussion on the other two susceptibilities to Sec.~\ref{sec:lensing}. 

The THz 2DCS typically measures the nonlinear response in the time domain~\cite{Woerner2013}. Following the discussion in Sec.~\ref{sec:intro}, we consider a three-pulse set up (Fig.~\ref{fig:schematic}): three circularly polarized electromagnetic pulses arrive at the sample successively at time $0$, $\tau$, $\tau+t_\mathrm{w}$, where $\tau,t_\mathrm{w}>0$ are the delay between successive pulses. The signal from the sample detected at a later time $t>0$ after the last pulse contains contributions from both linear and nonlinear responses. Rerunning the experiment with individual pulses, one may subtract off the linear response and thereby isolate the nonlinear response. 

The nonlinear signal is a convolution of the pulse profile and the third-order magnetic susceptibility:
\begin{multline}
\chi^{(3)}_{+--+} (t,t_\mathrm{w},\tau) =  \iiint^\infty_{-\infty} dx_1dx_2dx_3 
\\
 \widetilde{\chi}^{(3)}_{+--+} (t,x_1;\, t+t_\mathrm{w},x_2;\, t+t_\mathrm{w}+\tau,x_3).
\label{eq:signal}
\end{multline}
Here, $\chi^{(3)}_{+--+}$ is the optical susceptibility that depends only on time, while $\widetilde{\chi}^{(3)}_{+--+}$ is the  \emph{spacetime-dependent} susceptibility. Note the time parametrization of the latter quantity follows the standard convention for nonlinear susceptibility~\cite{Boyd2008}, whereas the former does not. We use the symbol with and without the tilde to emphasize these differences.

We visualize $\chi^{(3)}_{+--+}(t,t_\mathrm{w},\tau)$ by holding $t_\mathrm{w}$ constant and scanning $\tau$ and $t$. We obtain the two-dimensional spectra $\chi^{(3)}(\omega_t,t_\mathrm{w},\omega_\tau)$ by performing a two-dimensional one-sided Fourier transform of $\chi^{(3)}(t,t_\mathrm{w},\tau)$ over the domain $t>0$ and $\tau>0$. Note alternative protocols for visualizing the nonlinear response exist~\cite{Woerner2013,Mahmood2020}. Ours is closely related to that of Refs.~\onlinecite{Lu2017,Wan2019}.

\section{Bosonization \label{sec:bosonization}}

In this section, we compute the nonlinear response of Eq.~\eqref{eq:xxz_hamil} by using bosonization. We show that the nonlinear susceptibility $\chi^{(3)}_{+--+}$ exhibits photon echo and characterize its features.

\subsection{Bosonization essentials \label{sec:essential}}

The bosonization of $S=1/2$ XXZ chain is standard~\cite{Affleck1988,Giamarchi2003}. We briefly review the results to establish notations. Upon bosonization, the Hamiltonian at the RG fixed point reads:
\begin{align}
H = \frac{u}{2\pi}\int(\frac{1}{K}(\nabla\phi)^2+K(\nabla\theta)^2)dx.
\label{eq:boson_hamil}
\end{align}
Here, $u$ is the speed of sound. $K$ is the Luttinger parameter. They may be computed from the microscopic model parameters $J_\perp,J_z,B$ by using the Bethe ans\"{a}tz~\cite{LutherPeschel1975,Haldane1980}. $\phi$ and $\theta$ are boson fields with the compactification conditions:
\begin{align}
\phi\sim \phi+\pi, \quad \theta \sim \theta+2\pi .
\label{eq:boson_compact}
\end{align}
They obey the non-local commutation relation:
\begin{align}
[\phi(x),\theta(y)] = -i\pi \Theta(x-y),
\end{align}
where $\Theta(\cdot)$ is the Heaviside step function. Note there is freedom in choosing the commutation relation between $\phi$ and $\theta$. We discuss the difference between the different commutation relation prescriptions and the associated subtleties in Appendix~\ref{app:prescription}.

Up to a cutoff dependent prefactor, the spin operators assume the following form:
\begin{align}
S^-_j \approx \left\{\begin{array}{cc}
\exp(i\theta(x)) & (J_\perp<0)\\
\exp(i\theta(x))\cos(2\phi(x)-2\pi mx) & (J_\perp>0)\\
\end{array}\right. ,
\label{eq:spin_op}
\end{align}
where $m$ is the magnetization density, and $x$ is the spatial coordinate of site $j$. The external field $B$ is subsumed into the expression of spin operator through $m$. 

Note we have omitted in Eq.~\eqref{eq:spin_op} the spatially staggered ($\propto (-)^j$) component, which does not contribute to the optical response as the wavelength of the THz probe is typically much larger than the lattice spacing. Although the ferromagnetic chain ($J_\perp <0$) and the antiferromagnetic chain ($J_\perp > 0$) are exactly mapped to each other by a $\pi$-rotation of spins about the $z$-axis at every other site, i.e. $S^{\pm}_j \to (-1)^j S^{\pm}_j$, such mapping also exchanges the uniform and staggered components of the spin operator. Since we only consider the uniform component, it is necessary to distinguish the two cases for our purpose as shown in Eq.~\eqref{eq:spin_op}.

\subsection{Four-point response function \label{sec:gr}}

The experimentally measured signal is related to the spacetime-dependent nonlinear magnetic susceptibility $\widetilde{\chi}^{(3)}_{+--+}$ (Eq.~\eqref{eq:signal}).
Its Kubo formula~\cite{Kubo1957} reads: 
\begin{multline}
\widetilde{\chi}^{(3)}_{+--+}(1,2,3) =  i^3\Theta(t_1)\Theta(t_2-t_1)\Theta(t_3-t_2) \\
\times \langle [[[S^+(0),S^-(-1)],S^-(-2)],S^+(-3)] \rangle.
\label{eq:kubo}
\end{multline}
$0,1,2,3$ are shorthand notations for spacetime coordinates $(0,0)$, $(t_1,x_1)$, $(t_2,x_2)$, and $(t_3,x_3)$, respectively. $\widetilde{\chi}^{(3)}_{+--+}$ measures the system's response at the spacetime origin due to successive perturbations at $(-t_3,-x_3)$, $(-t_2,-x_2)$, and $(-t_1,-x_1)$. 

From Eq.~\eqref{eq:spin_op}, we see that the spin operators are linear combinations of vertex operators. It will be convenient to seek a general expression for the four-point response function with the following form:
\begin{align}
G^R(1,2,3) =   i^3\Theta(t_1)\Theta(t_2-t_1)\Theta(t_3-t_2) \nonumber\\
\times \langle [[[V_0(0),V_1(-1)],V_2(-2)],V_3(-3)] \rangle,
\label{eq:gr_def}
\end{align}
where
\begin{align}
V_j = e^{i(m_j\theta+2n_j\phi)},
\end{align}
are the vertex operators. We focus on the \emph{local} vertex operators, i.e. those preserve the boson compactification conditions~\eqref{eq:boson_compact}. This imposes the condition on the coefficients
\begin{align}
m_j,n_j\in\mathbb{Z} .
\label{eq:vertex_quant}
\end{align}
We also impose the charge neutrality condition $\sum_j m_j = \sum_j n_j = 0$ to ensure $G^R$ does not vanish in the thermodynamic limit. 

Eq.~\eqref{eq:gr_def} can be calculated by using the established technique~\cite{Giamarchi2003}. Here we only sketch the key steps. Using the Baker-Campbell-Hausdorff formula, we find the commutator $[e^{iA},e^{iB}] = -2\mathrm{sinh}\frac{[A,B]}{2}\,e^{i(A+B)}$, where $A,B$ are arbitrary linear combinations of $\phi$ and $\theta$. This permits a straightforward evaluation of the nested commutators in Eq.~\eqref{eq:gr_def}. We then compute the thermal average by using $\langle \exp(iA)\rangle = \exp(-\langle A^2\rangle/2)$, where $A$ is an arbitrary linear combination of $\theta$ and $\phi$. We obtain:
\begin{multline}
G^R(1,2,3) = -8\Theta(t_1)\Theta(t_2-t_1)\Theta(t_3-t_2)
\\
\times \sin(\alpha_{10})\sin(\alpha_{20}+\alpha_{21})\sin(\alpha_{30}+\alpha_{31}+\alpha_{32})
\\
\times C_{10}C_{20}C_{21}C_{30}C_{31}C_{32}.
\label{eq:gr_result}
\end{multline}

The above is the main result of this subsection.  Here, $\alpha_{ij}$ and $C_{ij}$ are defined for the vertex operator $V_i$ and $V_j$. $\alpha_{ij}$ comes from the commutator of vertex operators:
\begin{subequations}
\begin{multline}
\alpha_{ij} = \frac{\pi}{2}  (l_i l_j \mathrm{Sgn}(x^+_{ij})+r_i r_j\mathrm{Sgn}(x^-_{ij})
\\ 
 + l_ir_j-r_il_j).
\end{multline}
We have used light cone coordinates $x^\pm \equiv ut \pm x $. $x^\pm_{ij} \equiv x^\pm_i -x^\pm_j$. $\mathrm{Sgn}(\cdot)$ is the sign function. $l_j$ and $r_j$ are real parameters related to $m_j,n_j$ through:
\begin{align}
l_j = \frac{m_j}{2\sqrt{K}}+\sqrt{K}n_j;\quad
r_j = \frac{m_j}{2\sqrt{K}}-\sqrt{K}n_j.
\end{align}
Meanwhile,
\begin{align}
C_{ij} = \left|\frac{\sinh(\pi Tx^+_{ij}/u)}{\pi T\epsilon /u}\right|^{l_il_j}\left|\frac{\sinh(\pi Tx^-_{ij}/u)}{\pi T\epsilon /u}\right|^{r_ir_j},
\label{eq:C_ij}
\end{align}
\end{subequations}
where $\epsilon$ is the short-distance cutoff. $T$ is the temperature. 

Causality is an important property shared by all experimentally accessible response functions. For a relativistic system described by the fixed point Hamiltonian Eq.~\eqref{eq:boson_hamil}, the response vanishes if the perturbations are outside the past light cone of the detection event. It is then natural to ask if Eq.~\eqref{eq:gr_result} is causal and under what conditions. In Appendix~\ref{app:causality}, we show that Eq.~\eqref{eq:gr_result} is causal provided that the vertex operators are \emph{local}, i.e. Eq.~\eqref{eq:vertex_quant} holds for all $V_j$. It is straightforward to check that the vertex operators that appear in the expression of $S^\pm(x)$ (Eq.~\eqref{eq:spin_op}) indeed fulfill this condition. Thus, $\widetilde{\chi}^{(3)}_{+--+}$ is causal as expected.

\subsection{FM chain \label{sec:boson_FM}}

\begin{figure}
\includegraphics[width = \columnwidth]{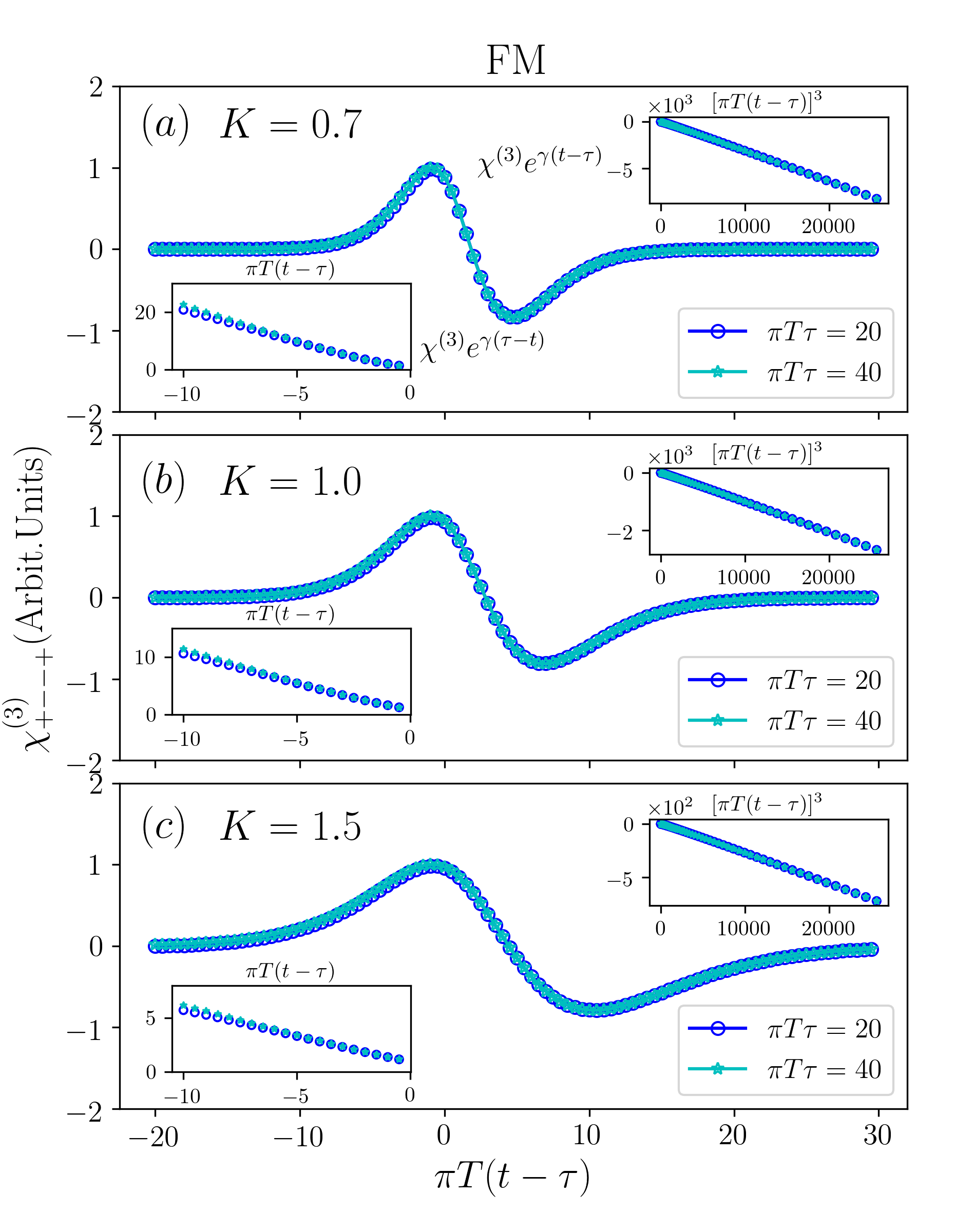}
\caption{(a) Nonlinear magnetic susceptibility $\chi^{(3)}_{+--+}$ of a ferromagnetic chain as a function of $\pi T(t-\tau)$ at $\pi T\tau = 20$ (blue) and $\pi T\tau = 40$ (cyan). The waiting time $\pi T t_\mathrm{w} = 1$. Luttinger parameter $K=0.7$. Data for different $\tau$ are scaled by the \emph{same} factor such that the maximum of $\pi T\tau = 20$ data is 1. Upper right inset: the $t>\tau$ part of the data, multiplied by $\exp(\gamma|t-\tau|)$ with $\gamma = \pi T/(2K)$, plotted against $(\pi T(t-\tau))^3$. Lower left inset: the $t<\tau$ part of the data, multiplied by $\exp(\gamma|t-\tau|)$, plotted against $\pi T(t-\tau)$. (b) The same as (a) but for $K=1$. (c) The same as (a) but for $K=1.5$.
}
\label{fig:asym_fm}
\end{figure}

\begin{figure}
\includegraphics[width = \columnwidth]{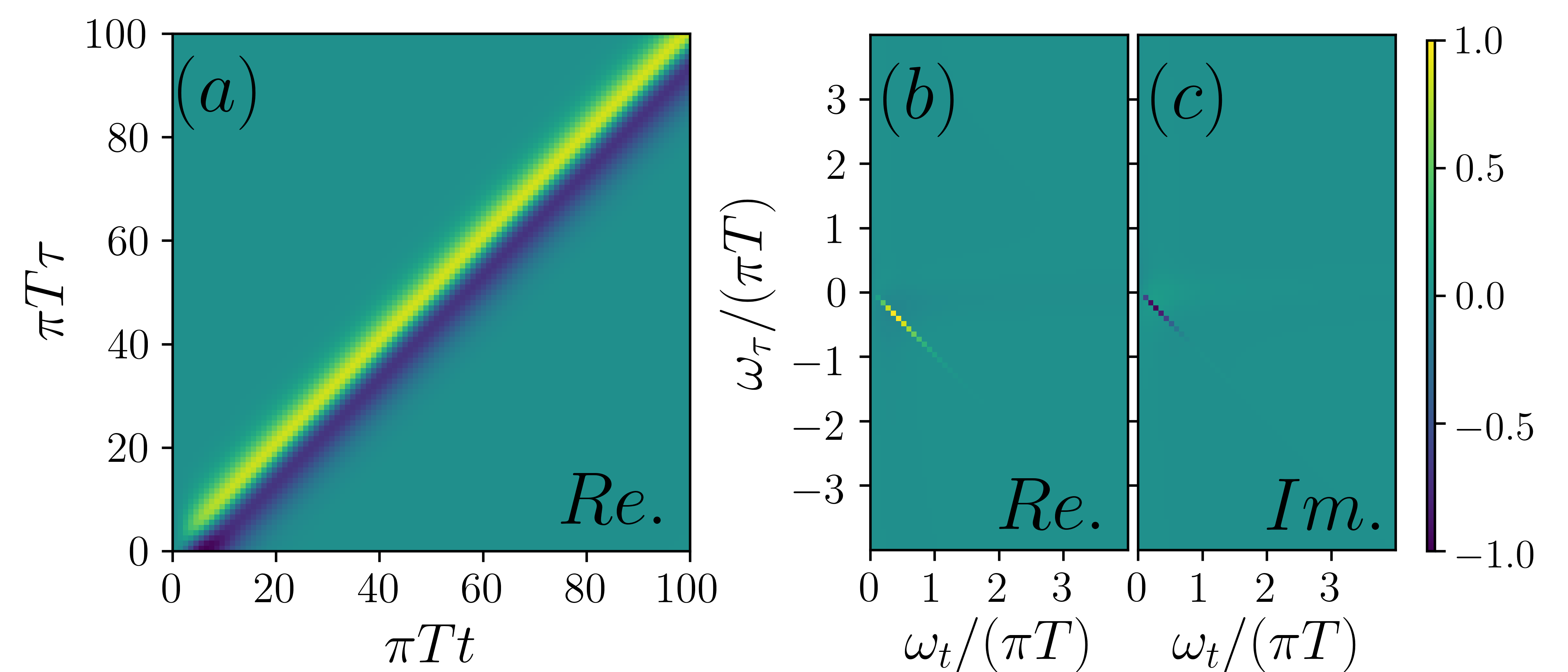}
\caption{(a) Nonlinear magnetic susceptibility $\chi^{(3)}_{+--+}$ of a ferromagnetic chain as function of $\pi T t$ and $\pi T\tau$. The waiting time $\pi T t_\mathrm{w} = 1$. The Luttinger parameter $K=1$. The data are real numbers and have been rescaled such that the maximum is 1. (b)(c) The real and imaginary parts of the two-dimensional spectrum, obtained by Fourier transforming the data of (a). Only the first and fourth quadrants are shown. The data in the other two quadrants are obtained by complex conjugation, i.e. the real part (b) is invariant under the transformation $\omega_\tau\to-\omega_\tau, \omega_t\to-\omega_t$, whereas the imaginary part (c) changes sign.}
\label{fig:ana_fm}
\end{figure}

In this subsection, we consider the ferromagnetic ($J_\perp<0$) chain. Plugging Eq.~\eqref{eq:spin_op} into the Kubo formula (Eq.~\eqref{eq:kubo}) yields:
\begin{multline}
\widetilde{\chi}^{(3)}_{+--+}(1,2,3) \overset{(J_\perp<0)}{=}  i^3\Theta(t_1)\Theta(t_2-t_1)\Theta(t_3-t_2) 
\\
\times \langle [[[e^{-i\theta(0)},e^{i\theta(-1)}],e^{i\theta(-2)}],e^{-i\theta(-3)}] \rangle.
\label{eq:boson_fm_chi3}
\end{multline}
The above has the form of Eq.~\eqref{eq:gr_def}. We may read off its explicit expression from Eq.~\eqref{eq:gr_result} by setting $l = r =1/\sqrt{4K}$ for $\exp(i\theta)$ and $l=r=-1/\sqrt{4K}$ for $\exp(-i\theta)$. An immediate consequence of Eq.~\eqref{eq:gr_result} is that $\chi^{(3)}_{+--+}$ is strictly real and independent of the magnetization density $m$. 

The next step is to find $\chi^{(3)}_{+--+}$ by integrating $\widetilde{\chi}^{(3)}_{+--+}$ over spatial coordinates (Eq.~\eqref{eq:signal}). Given the complex structure of the integrand, the integral is unlikely to admit a simple, closed form.  We instead seek the asymptotic form of $\chi^{(3)}_{+--+}$ valid when $t,t_w,\tau$ are large. 

To this end, we use the following approximation for the function $C_{ij}$ (Eq.~\eqref{eq:C_ij}):
\begin{align}
C_{ij}(x_{ij},t_{ij}) &\approx \exp\left( \frac{\pi T}{u}(l_il_j|x^+_{ij}|+r_ir_j|x^-_{ij}|) \right).
\label{eq:c_ij_apprx}
\end{align}
We have omitted a cutoff dependent prefactor. Eq.~\eqref{eq:c_ij_apprx} captures the exponential falling-off/growth of $C_{ij}$ away from the light cone but neglects the algebraic singularity near the light cone $x_{ij}=\pm ut_{ij}$. The latter is short distance physics and should not affect the asymptotic behavior. The validity of Eq.~\ref{eq:c_ij_apprx} will be further verified \emph{a posteriori} by numerical integration.

With the approximation Eq.~\eqref{eq:c_ij_apprx}, the integral Eq.~\eqref{eq:signal} is now elementary. After lengthy calculations, we find the integration produces two groups of terms: The first group of terms simply decrease as $t$ or $\tau$ increases, which we discard as they are uninteresting for our purpose. The second group of terms show signature of photon echo --- as we fix $\tau$ and scan $t$, they exhibit a maximum near $t\approx \tau$. Retaining these terms, we find:
\begin{align}
\chi^{(3)}_{+--+}  \overset{(J_\perp<0)}{\sim}  \left\{ \begin{array}{cc}
(\tau-t)\,e^{-\frac{\pi T(\tau-t)}{2K}} & (t\ll \tau) \\
-(t-\tau)^3 e^{-\frac{\pi T(t-\tau)}{2K}} & (t\gg\tau) 
\end{array}\right. .
\label{eq:chi3_fm_apprx}
\end{align}
The above is the key result of this subsection.

Eq.~\eqref{eq:chi3_fm_apprx} suggests that the photon echo is perfect, i.e. it is a function of $t-\tau$. The time scale of the photon echo signal is set by $2K/(\pi T)$. Moreover, it is independent of the waiting time $t_\mathrm{w}$ in the asymptotic regime. As we have discussed in Sec.~\ref{sec:intro}, the photon echo in a few-body system is understood as the result of the rephasing process, which in turn builds on transitions between discrete energy levels. Since the energy spectrum of the Luttinger spin liquid is continuous, the rephasing picture does not apply. The physics behind the photon echo in the present system will be discussed in detail in Sec.~\ref{sec:lensing}.

We test the validity of Eq.~\eqref{eq:chi3_fm_apprx} by performing the integration Eq.~\eqref{eq:signal} numerically. We use a short distance cutoff  $\pi \epsilon T/u = 1$, which smoothens the algebraic singularities and discontinuities that would otherwise appear in the integrand in the limit $\epsilon\to0^+$. With finite $\epsilon$, the integrand decreases rapidly outside the light cone of $(0,0)$. We thus limit the domain of integration to the light cone plus a small interval of size $R$ beyond the light cone. We set $\pi TR/u = 2$ with the relative error $O(10^{-3})$.

Fig.~\ref{fig:asym_fm} shows $\chi^{(3)}_{+--+}$ for representative Luttinger parameters $K=0.7$, $K=1$, and $K=1.5$. $\pi T t_\mathrm{w} = 1$ and $\pi T \tau=20,40$. For all cases, the nonlinear response shows a surge near $t\approx \tau$, exhibiting the clear signature of photon echo. Note the maximum is \emph{not} exactly located at $t=\tau$ but fairly close to it. The data for different values of $\tau$ overlay within numerical error when plotted as a function of $t-\tau$. This demonstrates the photon echo is independent of $\tau$ for large $\tau$. 

Numerical integration also indicates that the value of $\chi^{(3)}_{+--+}$ measured at $t=\tau$ does not depend on $t_\mathrm{w}$ within numerical error in the asymptotic regime $\pi T\tau\gg 1$, which is in agreement with Eq.~\eqref{eq:chi3_fm_apprx}

We further examine the asymptotic behavior of $\chi^{(3)}_{+--+}$ by multiplying it with $\exp(\pi T |t-\tau|/(2K))$. Eq.~\eqref{eq:chi3_fm_apprx} suggests the product would be $\propto \tau-t$ when $t<\tau$, and $\propto -(t-\tau)^3$  when $t>\tau$. The insets of Fig.~\ref{fig:asym_fm} show that its behavior is in excellent agreement with Eq.~\eqref{eq:chi3_fm_apprx}. 

Having analyzed the photon echo signal for fixed value of $\tau$, we now scan $\tau$ and present the nonlinear response as a function of both $\tau$ and $t$. Fig.~\ref{fig:ana_fm}(a) shows $\chi^{(3)}_{+--+}$ for $K=1$ and $\pi Tt_\mathrm{w} = 1$. The photon echo appears as a bright feature at the diagonal of the $(t,\tau)$ plane. This feature persists along the diagonal direction, highlighting the fact that the photon echo is perfect. 

Performing the FFT of the time domain data yields the two-dimensional spectrum (Fig.~\ref{fig:ana_fm}(b)\&(c)). In the frequency domain, the photon echo manifest itself as a pair of highly anisotropic peaks in the second/fourth quadrants, symmetrically distributed with respect to the origin.  The peak width along the diagonal direction of the second/fourth quadrant scales with $T$. The peak width along the anti-diagonal direction is resolution limited --- the photon echo signal in the time domain (Fig.~\ref{fig:ana_fm}(a)) is independent of $t+\tau$ at late time, and hence its Fourier transform $\sim \delta(\omega_t+\omega_\tau)$. 

To recapitulate, the $\chi^{(3)}_{+--+}$ of the ferromagnetic chain is real and independent of the magnetization density. It exhibits perfect photon echo, which depends on $t-\tau$ rather than $t$ and $\tau$ both.

\subsection{AFM chain \label{sec:boson_AFM}}

We turn to the antiferromagnetic ($J_\perp>0$) case in this subsection. We write the spin operator as:
\begin{align}
S^-(x) & = e^{-2\pi imx}e^{i(\theta(x)+2\phi(x))}+ e^{2\pi imx}e^{i(\theta(x)-2\phi(x))}
\nonumber\\
& =  e^{-2\pi imx} a(x) + e^{2\pi imx}b(x),\quad (J_\perp>0)
\end{align}
where we have defined vertex operators $a$ and $b$ for later convenience. Inserting the above into Eq.~\eqref{eq:kubo} yields:
\begin{widetext}
\begin{align}
\widetilde{\chi}^{(3)}_{+--+}(1,2,3) \overset{(J_\perp>0)}{=}  e^{2\pi im (x_1+x_2-x_3)}G^R_{a^\dagger aa a^\dagger}(1,2,3)  &+ e^{- 2\pi i m (x_1+x_2-x_3)}G^R_{b^\dagger bbb^\dagger}(1,2,3)
\nonumber \\
+ e^{2\pi im(x_1-x_2+x_3)} G^R_{a^\dagger a b b^\dagger}(1,2,3) & + e^{-2\pi im(x_1-x_2+x_3)} G^R_{b^\dagger b a a^\dagger}(1,2,3)
\nonumber \\
+ e^{2\pi im(x_1-x_2-x_3)} G^R_{b^\dagger a b a^\dagger}(1,2,3) & + e^{-2\pi im(x_1-x_2-x_3)} G^R_{a^\dagger b a b^\dagger}(1,2,3).
\label{eq:boson_afm_chi3}
\end{align}
We have defined a set of response functions with the form of Eq.~\eqref{eq:gr_def}. For instance, $G^R_{a^\dagger a a a^\dagger}$ is defined as:
\begin{align}
G^R_{a^\dagger aaa^\dagger}(1,2,3) =   i^3\theta(t_1)\theta(t_2-t_1)\theta(t_3-t_2)  \langle [[[a^\dagger (0),a(-1)],a(-2)],a^\dagger (-3)] \rangle.
\end{align}
\end{widetext}
The other response functions are defined in the same vein. The expression for these response function can be read off from Eq.~\eqref{eq:gr_result} by plugging in appropriate values of $l_{1,2,3}$ and $r_{1,2,3}$: $l = \sqrt{K}+1/(2\sqrt{K})$ and $r = -\sqrt{K}+1/(2\sqrt{K})$ for vertex operator $a$; for vertex operator $b$, the value of $l$ and $r$ are switched. We have dropped the response functions that violate the charge neutrality condition (e.g. $G^R_{a^\dagger aab^\dagger}$) as they vanish in the thermodynamic limit. 

The calculation of $\chi^{(3)}_{+--+}$ parallels the ferromagnetic case (Sec.~\ref{sec:boson_FM}). Note, however, $\chi^{(3)}_{+--+}$ is now complex and depends on the magnetic field through the magnetization density $m$. We find only $G_{a^\dagger a a a^\dagger}$ and $G_{b^\dagger bb b^\dagger}$ contribute to photon echo. Using the approximation (Eq.~\eqref{eq:c_ij_apprx}), we obtain after lengthy calculation:
\begin{align}
\chi^{(3)}_{+--+} &\overset{(J_\perp>0)}{\sim} e^{-2\pi imu(t-\tau)}
\nonumber\\
&\times \left\{\begin{array}{cc}
e^{-\pi (\Delta-2) T(\tau-t)} & (t\ll \tau)\\
(t-\tau) e^{-\pi (\Delta-2) T(t-\tau)} & (t\gg \tau)
\end{array}\right. .
\label{eq:chi3_afm_apprx}
\end{align}
Here, we have defined a parameter $\Delta = 2K+1/(2K)$. The above is the key result of this subsection.

Eq.~\eqref{eq:chi3_afm_apprx} shows that the $\chi^{(3)}_{+--+}$ of antiferromagnetic chain exhibits perfect photon echo similar to the ferromagnetic chain. Different from the ferromagnetic chain, the signal now shows oscillations with the frequency set by the magnetization density $m$, which, in turn, depends on the magnetic field $B$. The time scale of the signal is $1/((\Delta-2)\pi T)$. In particular, in the Heisenberg limit where $K\to 1/2$ ($\Delta\to 2$), the time scale diverges --- this reflects the Larmor precession of the total magnetization, which we elaborate on in Appendix~\ref{app:heisenberg}.

\begin{figure}
\includegraphics[width = \columnwidth]{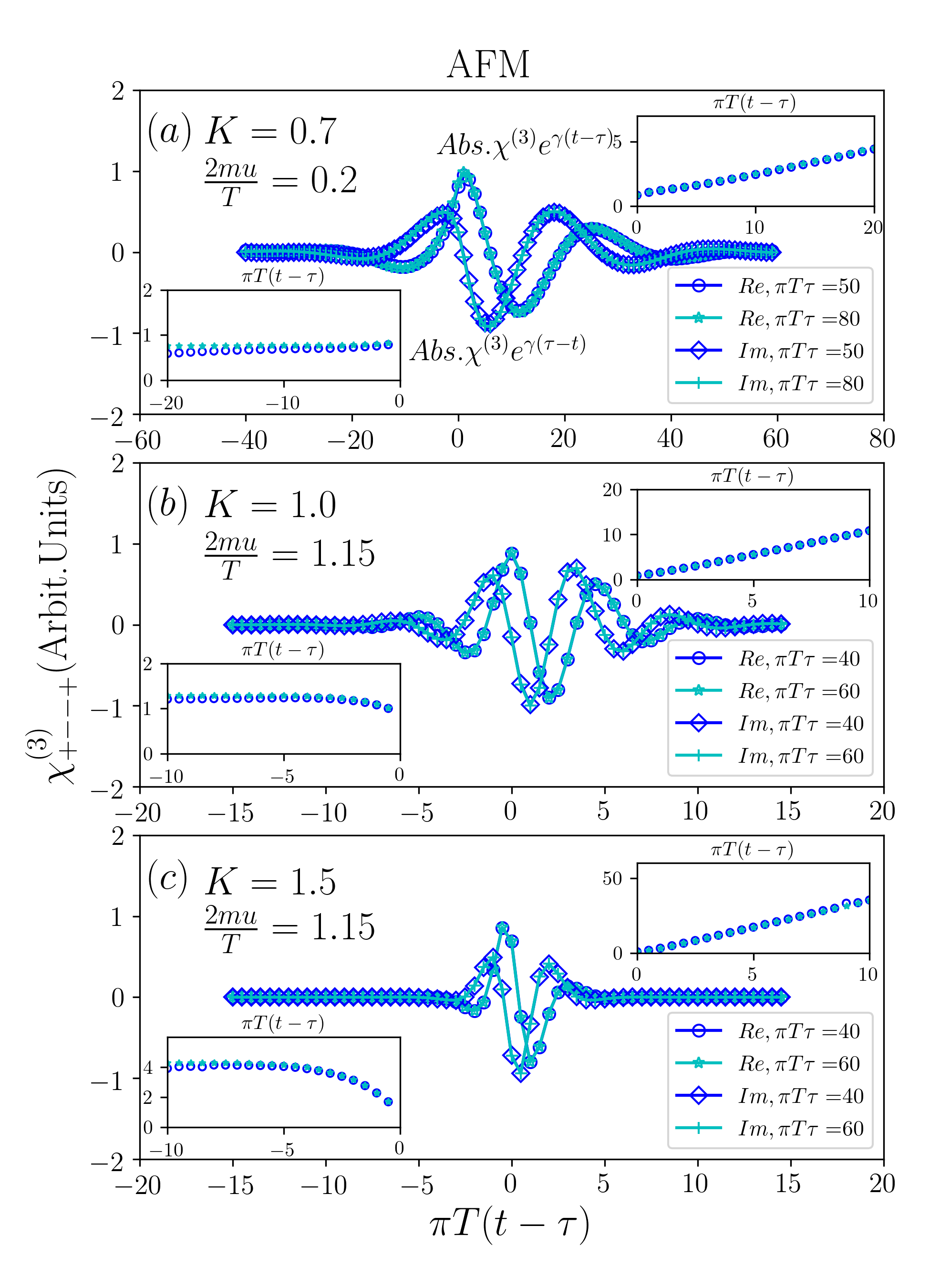}
\caption{(a) Nonlinear magnetic susceptibility $\chi^{(3)}_{+--+}$ of an antiferromagnetic chain as a function of $\pi T(t-\tau)$ for $\pi T\tau = 50$ (blue) and $\pi T\tau = 80$ (cyan). The waiting time $\pi T t_\mathrm{w} =1 $. Luttinger parameter $K=0.7$. The data for different $\tau$ are rescaled by the \emph{same} factor such that the maximum of the complex modulus of the $\pi T\tau=50$ data is 1. Upper right inset shows the complex modulus of the $t>\tau$ part of the data, multiplied by $\exp(\gamma|t-\tau|)$ with $\gamma = \pi (\Delta-2) T$. Lower left inset the complex modulus of the $t<\tau$ part of the data, multiplied by $\exp(\gamma|t-\tau|)$ (b) The same as (a) but for $K=1$. (c) The same as (a) but for $K=1.5$}
\label{fig:asym_afm}
\end{figure}

We assess the validity of Eq.~\eqref{eq:chi3_afm_apprx} by numerical integration. Fig.~\ref{fig:asym_afm} show $\chi^{(3)}_{+--+}$ for representative Luttinger parameters $K=0.7$, $K=1$, and $K=1.5$. $\pi T t_\mathrm{w}=1$. The echo is clearly visible from the data. The data for different values of $\tau$ overlay when plotted as a function of $t-\tau$, demonstrating that the echo is perfect. Numerical integration suggests $\chi^{(3)}_{+--+}$ is independent of $t_\mathrm{w}$ within numerical error when $t$ and $\tau$ are large.

To test the asymptotic behavior of $\chi^{(3)}_{+--+}$, we multiply its complex modulus with $\exp(\pi(\Delta-2)|t-\tau|)$. Eq.~\eqref{eq:chi3_afm_apprx} suggests the product shows linear behavior for $t>\tau$ and approaches a constant for $t<\tau$. The insets of Fig.~\ref{fig:asym_afm} show good agreement with Eq.~\eqref{eq:chi3_afm_apprx}. For $K=0.7$ and $\pi T\tau = 50$, the data deviate slightly from the constant behavior; we think this occurs because $\tau$ is not sufficiently large to suppress the non-asymptotic contributions.

\begin{figure}
\includegraphics[width = \columnwidth]{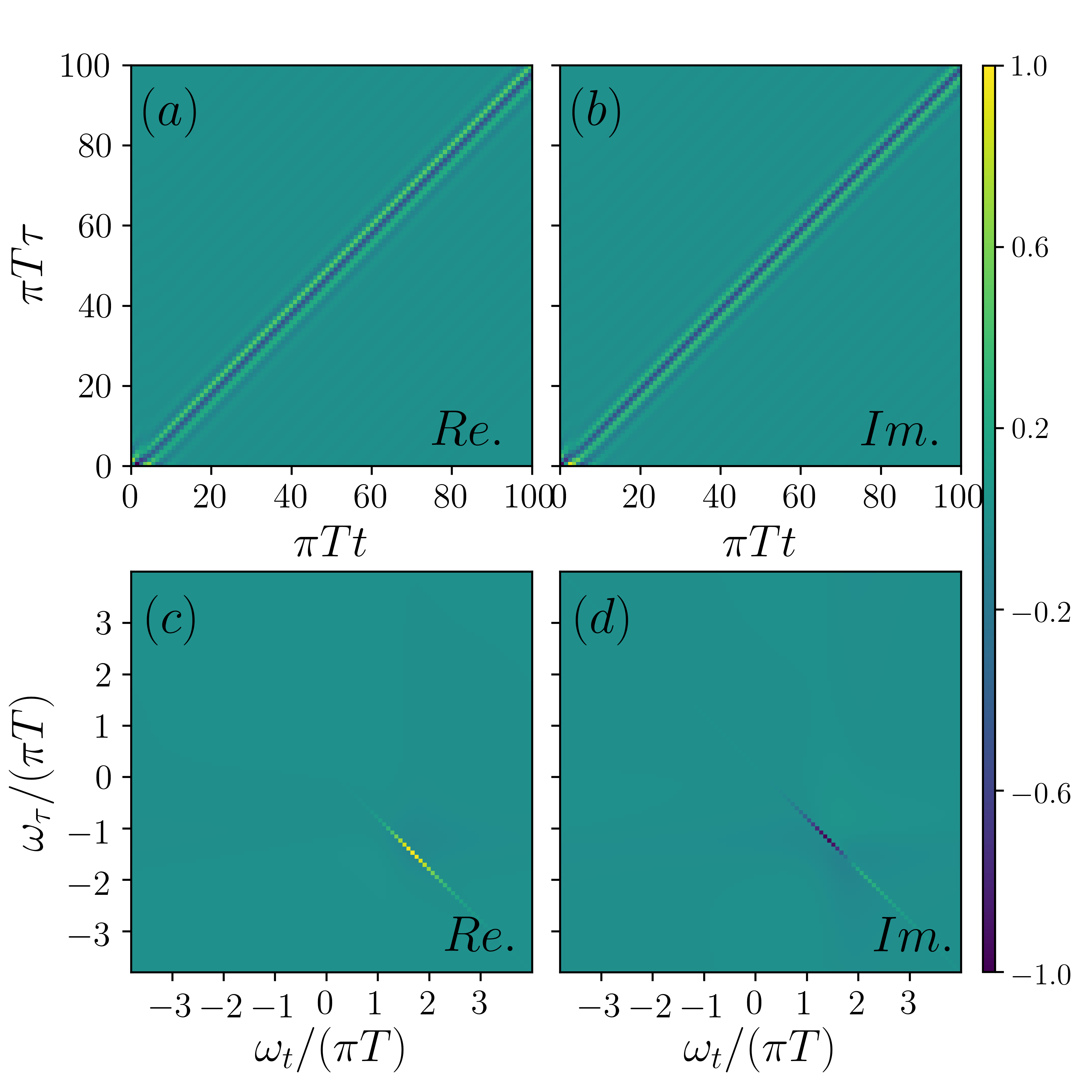}
\caption{(a)(b) The real and imaginary parts of the nonlinear magnetic susceptibility $\chi^{(3)}_{+--+}$ of an antiferromagnetic chain as function of $\pi Tt$ and $\pi T\tau$. The Luttinger parameter $K=1$. Magnetization density $2mu/T = 1.15$. The waiting time $\pi Tt_\mathrm{w}=1$. The data are scaled such that the maximum of the absolute value of the data is 1. (c)(d) The real and imaginary parts of the two-dimensional spectrum.}
\label{fig:ana_afm}
\end{figure}

We then present the dependence of the photon echo signal on both $\tau$ and $t$. Fig.~\ref{fig:ana_afm}(a)\&(b) show the real and imaginary parts of $\chi^{(3)}_{+--+}$ as a function of $t$ and $\tau$ for the Luttinger parameter $K=1$. The magnetization density $2mu/T = 1.15$. The waiting time $\pi Tt_\mathrm{w}=1$. The photon echo appears as the bright feature persists along the diagonal direction ($t=\tau$). Performing Fourier transform over $t$ and $\tau$ produces the two-dimensional spectrum (Fig.~\ref{fig:ana_afm}(c)\&(d)). The photon echo appears in the frequency domain as a highly anisotropic peak in the fourth quadrant. The peak is approximately located at $\omega_t = 2\pi mu$ and $\omega_\tau = -2\pi mu$ as suggested by Eq.~\eqref{eq:chi3_afm_apprx}. Its width along the diagonal of the fourth quadrant scales with $T$, whereas its width along the anti-diagonal direction is resolution limited.
 
To recapitulate, the $\chi^{(3)}_{+--+}$ of the antiferromagnetic chain shows clear signature of perfect photon echo. The echo signal is oscillatory with the frequency set by magnetization density.

\section{Spinon Lensing Picture \label{sec:lensing}}

\begin{figure}
\centering
\includegraphics[width=\columnwidth]{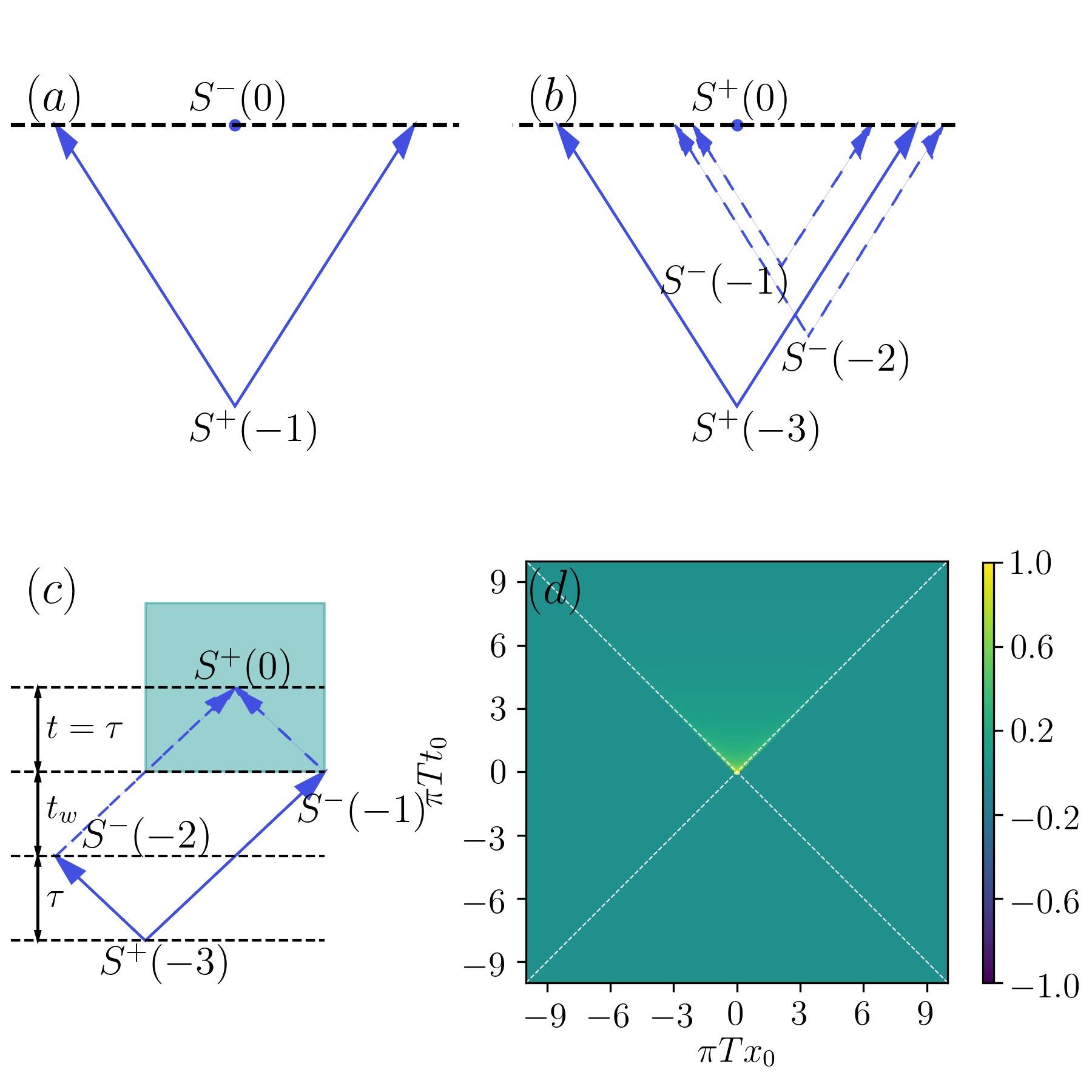}
\caption{(a) The two-point correlation function (Eq.~\eqref{eq:2point_vertex}) can be understood in terms of a spinon creation/annihilation process. The spin raising operator at $(-t_1,-x_1)$ creates a pair of spinons with left and right chirality, which then propagate with velocities $\pm u$ (blue solid lines). The spin lowering operator tries to annihilate the spinon pair at $(0,0)$. (b) The four-point correlation function Eq.~\eqref{eq:R1} corresponds to the spinon creation at $(-t_3,-x_3)$ and $(0,0)$ (blue solid lines), and antispinon creation at  $(-t_2,-x_2)$ and $(-t_3,-x_3)$ (blue dashed lines). (c) In the lensing configuration (Eq.~\eqref{eq:lensing_config}), the left-moving spinon created at $(-t_3,-x_3)$ is captured by the spin lowering operator at $(-t_2,-x_2)$ and converted to a right-moving antispinon. Likewise, the right-moving spinon is captured by the spin lowering operator at $(-t_1,-x_1)$ and converted to a left-moving antispinon. The antispinons then meet at $(0,0)$ and are annihilated by the spin raising operator. (d) Behavior of $\widetilde{\chi}^{(3)}_{+--+}$ in the shaded region in (c), which shows its magnitude reaches the maximum near the ``focal point" $(0,0)$. Here, $\pi Tt = \pi T\tau = 10$, and $\pi T t_\mathrm{w}= 10$. 
}
\label{fig:lensing}
\end{figure}

In the previous section, we have shown that the nonlinear magnetic susceptibility $\chi^{(3)}_{+--+}$ of the Luttinger spin liquid exhibits photon echo that resembles the perfectly rephasing echo in a few-body system such as a single spin. In this section, we provide an intuitive picture that clarifies its origin in the many-body system under consideration, namely Luttinger spin liquids. We illustrate our picture on the ferromagnetic spin chain and then generalize it to the antiferromagnetic chain, and discuss its various features and implications.

To set the stage, we review the effect of the bosonized spin raising operator $\exp(-i\theta)$. Let us consider the time evolution after the operator $\exp(-i\theta(0,0))$ at $t=x=0$ on the initial state $|\Psi(0) \rangle$ at $t=0$ which is assumed to be an energy eigenstate. At time $t$, the state is given by
\begin{align} 
|\Psi(t) \rangle &=  e^{-i \mathcal{H} t } e^{-i\theta(0,0)} | \Psi(0) \rangle 
\nonumber \\
& =  e^{-i \mathcal{H} t } e^{-i\theta(-t,0)} e^{+i \mathcal{H} t } e^{-i \mathcal{H} t } | \Psi(0) \rangle 
\nonumber \\
& \sim e^{-i\theta(-t,0)} | \Psi(0) \rangle ,
\end{align}
where we used the assumption that $\Psi(0)\rangle$ is an energy eigenstate and dropped the overall phase factor.
We can decompose the field $\theta$ to chiral components as 
\begin{align}
    \theta(t,x) = & \theta_L(x^+) + \theta_R(x^-) ,
    \label{eq:chiral_decomp}
\end{align}
where
\begin{align}
\theta_L = \frac{1}{2} \left(\theta + \frac{\phi}{K} \right);
\quad
\theta_R = \frac{1}{2} \left(\theta - \frac{\phi}{K} \right).
\end{align}
The equation of motion implies that each chiral components depends only on the corresponding light cone coordinate,
as in Eq.~\eqref{eq:chiral_decomp}.
Using this, we can translate the time dependence into the position dependence, so that 
\begin{align} 
    |\Psi(t) \rangle \sim &  e^{-i\theta_L(-ut)} e^{-i\theta_R(ut)}  | \Psi(0) \rangle ,
    \label{eq:spinon_evol}
\end{align}
where $\theta_{L,R}(\mp ut)$ is now understood as a static operator at locations $\mp ut$,
up to the overall phase factor which includes the one coming from the commutation relation between $\theta_R$ and $\theta_L$.

Now, recall the equal-time commutation relation:
\begin{align}
\phi(x)  e^{-i\theta_{L,R}(y)} = e^{-i\theta_{L,R}(y)} \left(\phi(x) - \frac{\pi}{2} \Theta{(x-y)} \right) .
\end{align}
Namely, $\exp(-i\theta_{L,R}(y))$ creates a kink of step $\pi/2$ in $\phi$-field at the location $y$. Since this step corresponds to the magnetization density $(1/2)\,\delta(x-y)$, this kink can be interpreted as a spinon~\cite{Pham2000}. Eq.~\eqref{eq:spinon_evol} then implies that the application of the operator $\exp(-i\theta)$ at $t=0$ is equivalent to creation of a right-moving spinon at $x=+ut$ and a left-moving spinon at $x=-ut$ at time $t$, when the other operations are applied after the time $t$. The chiral vertex operators $\exp(-i\theta_{L,R})$ can be viewed as spinon creation operators of the corresponding chirality. This justifies the following simple visual picture: Applying $\exp(-i \theta(0,0))$ creates a pair of the right-moving and left-moving spinons at $t=x=0$. Then these spinons propagate with the constant velocity $\pm u$.

Real-time correlation functions of the vertex operators can be qualitatively understood in terms of this visual picture. As the simplest example, consider the two-point correlation function (Fig.~\ref{fig:lensing}a)
\begin{align}
& \langle S^-(0)S^+(-1)\rangle \sim \langle e^{i\theta(0,0)}e^{-i\theta(-t_1,-x_1)} \rangle 
\nonumber\\
&\quad\sim \langle e^{i\theta_L(0)} e^{i\theta_R(0)}  e^{-i\theta_L(-x_1-ut_1)} e^{-i\theta_R(-x_1+ut_1)}\rangle
\nonumber\\
& \quad \sim  \left|\frac{\sinh(\pi Tx^+_1)}{\pi T\epsilon}\right|^{-\frac{1}{4K}}\left|\frac{\sinh(\pi Tx^-_1)}{\pi T\epsilon}\right|^{-\frac{1}{4K}}.
\label{eq:2point_vertex}
\end{align}
Here, we set $t_1>0$ and omit an overall phase factor. This correlation function corresponds to the creation of a pair of spinons at $(-t_1,-x_1)$, and the annihilation of a pair of spinons (or equivalently the creation of a pair of anti-spinons) at the origin. From the equation of motion (the second line), this correlation function can be interpreted as an expectation value of spinon creation operators at the split locations $-x_1 \pm ut_1$ and \emph{two} spinon annihilation operators at $0$. For a generic choice of $t_1,x_1$, these three locations are different. As a result, the expectation value, although does not vanish completely, is exponentially suppressed with respect to the correlation length $u/T$ at finite temperature $T$. 

If we wish to maximize the correlation function, we should try to ``catch" the spinons created at $(-t_1,-x_1)$ at the later time and annihilate them. If we can manage to annihilate all the spinons created earlier, the correlation function would be reduced to the equal-time correlation function of the identity operator, which does not decay as a function of time $t$ at all! Unfortunately, it is impossible to annihilate both spinons created at $(-t_1,-x_1)$ by the single operator $\exp(i\theta(0,0))$, since spinons split and are located on different positions at later times. 

We note it is possible to annihilate one of them, for example the left-moving spinon, by choosing the location of the second operator on the left side of the light cone ($x_1 = -ut_1$, or $x^+_1 = 0$). Setting $x^+_1=0$ in Eq.~\eqref{eq:2point_vertex} maximizes the first factor, which is formally divergent in the scaling limit (the cutoff $\epsilon\to0^+$) but bounded to be a finite constant due to the finite $\epsilon$ in a realistic system. However, $x^+_1=0$ implies $x^-_1 = 2 u t_1$, leading to the exponential suppression of the other factor in Eq.~\eqref{eq:2point_vertex}. Thus the two-point correlation function is exponentially suppressed for large $t_1$, for any choice of the relative spatial location $x_1$.

The situation is quite different for four-point correlation functions and corresponding response functions. It is convenient to rewrite the response function Eq.~\eqref{eq:boson_fm_chi3} as a sum over various four-point correlation functions by expanding the nested commutators:
\begin{subequations}
\begin{align}
\widetilde{\chi}^{(3)}_{+--+}(1,2,3) &\sim R_1+R_2+R_3+R_4
\nonumber\\
& -R'_1-R'_2-R'_3-R'_4,
\end{align}
where
\begin{align}
R_1 &= \langle e^{-i\theta(0)} e^{i\theta(-1)} e^{i\theta(-2)} e^{-i\theta(-3)}\rangle;
\label{eq:R1}
\\
R_2 &= \langle e^{-i\theta(-3)} e^{i\theta(-1)} e^{-i\theta(0)} e^{i\theta(-2)}\rangle;
\\
R_3 &= \langle e^{-i\theta(-3)} e^{i\theta(-2)} e^{-i\theta(0)} e^{i\theta(-1)}\rangle;
\\
R_4 &=  \langle e^{i\theta(-2)} e^{i\theta(-1)} e^{-i\theta(0)} e^{-i\theta(-3)}\rangle.
\end{align}
\end{subequations}
$R'_i$ is related to $R_i$ by reversing the order of operators in the product. Following Ref.~\onlinecite{Mukamel1995,Hamm2011}, each contribution can be identified as a Liouville pathway.

We find that each four-point correlation function can be made free from exponential suppression for an appropriate choice of locations of the operators at given set of times. As exchanging the order of operators in $R_i$ only results in a phase factor, it is sufficient for our purpose to consider the correlation function $R_1$ (Fig.~\ref{fig:lensing}b). The spinon pairs are created at $(-t_3,-x_3)$ and $(0,0)$, and annihilated at $(-t_2,-x_2)$ and $(-t_1,-x_1)$. Unlike the case with the two-point correlation function (Eq.~\ref{eq:2point_vertex}), here, we can catch and annihilate all the spinons created earlier by choosing (Fig.~\ref{fig:lensing}c):
\begin{align}
x_1 = u\tau; \quad
x_2 = -u(\tau+t_\mathrm{w});\quad
x_3 = -ut_\mathrm{w} .
\label{eq:lensing_config}
\end{align}
For this configuration, the operator at $(-t_2,-x_2)$ annihilates the left-moving spinon created at $(-t_3,-x_3)$ and creates a right-moving antispinon. The operator at $(-t_1,-x_1)$ annihilates the right-moving spinon created at $(-t_3,-x_3)$ and creates a left-moving antispinon. The two antispinons moving to the opposite directions finally meet at the origin at time $0$, and are annihilated by the operator $\exp(i\theta)$! As a consequence, the magnitude of the four-point correlation function reaches its maximum and does not decay even in the long-time limit as long as the spatial coordinates are chosen according to Eq.~\eqref{eq:lensing_config}. We call this phenomenon, which was absent in the two-point correlation function, the \emph{spinon lensing}, that is, it is possible to focus the two (anti)spinons to the same point at time $0$, by placing the operators judiciously at earlier times. We also note that the spatial mirror reflection of Eq.~\eqref{eq:lensing_config} is equally valid for lensing. 

The above reasoning can be made precise by applying the equation motion to reduce the $4$-point correlation function to an equal-time correlation function at time $0$ of multiple operators --- spinon creation operators at locations $-x_3 \pm u t_3$ and two at $0$, and antispinon creation operators at locations $-x_2 \pm u t_2$ and $-x_1 \pm u t_1$ (Fig.~\ref{fig:lensing}b). Under the condition Eq.~\eqref{eq:lensing_config}, the locations of the spinon creation operators match with those of the spinon annihilation operators, and consequently the correlation function $R_1$ reaches the maximum.

The same phenomena occur in the other four-point correlation functions. Summing them up, the nonlinear response $\widetilde{\chi}^{(3)}_{+--+}$ as a function of spacetime coordinates $(t_1,x_1)$, $(t_2,x_2)$, and $(t_3,x_3)$, reaches a maximum at the lensing configuration Eq.~\eqref{eq:lensing_config}. Fig.~\ref{fig:lensing}d shows the behavior of $\chi^{(3)}_{+--+}$ as a function of the detection position in a representative lensing configuration. Here we choose $\pi T t = \pi T\tau = 10$ and $\pi Tt_\mathrm{w} = 10$. It shows that the response is maximum at the ``focal point" $(0,0)$. As we have used a finite short distance cutoff $\epsilon$ to regularize the algebraic singularity at the light cone, we have slightly shifted $x_1$ and $x_2$ away from the light cone by a small distance $\epsilon$ to suppress the effect of the regularizer. 

Now, the optical nonlinear response $\chi^{(3)}_{+--+}$ is related to $\widetilde{\chi}^{(3)}_{+--+}$ by spatial integration. When the time delays $t=\tau$, the spatial integration is dominated by the neighborhood of the lensing configuration Eq.~\eqref{eq:lensing_config} and its mirror reflection, and, as a result, remains non-vanishing even in the long-time limit $t=\tau\to\infty, t_\mathrm{w} \to \infty$. This explains the origin of the prefect photon echo observed in the $\chi^{(3)}_{+--+}$ of the ferromagnetic chain.

Turning to the antiferromagnetic chain, $\widetilde{\chi}^{(3)}_{+--+}$ is now a linear combination of various four-point response functions (Eq.~\eqref{eq:boson_afm_chi3}). Among these, $G^R_{a^\dagger a aa^\dagger}$ and $G^R_{b^\dagger bbb^\dagger}$ support the lensing phenonenon similar to Fig.~\ref{fig:lensing} with the only difference being the type of fractional excitations created/annihilated by the spin operators. In the antiferromagnetic chain, $S^+$ creates a pair of spinons plus a pair of Laughlin quasiparticle and quasihole~\cite{Pham2000}. Note the created spinon and the Laughlin quasiparticle (and similarly the created antispinon and the Laughlin quasihole) are superimposed on each other, and therefore the world lines shown in Fig.~\ref{fig:lensing} should be interpreted as the world line of the composite object. Our numerical calculations show that, indeed, only these two response functions produce photon echo whereas the others response functions do not. 

In Sec.~\ref{sec:setup}, we have pointed out that the nonlinear susceptibilities $\chi^{(3)}_{+-+-}$ and $\chi^{(3)}_{++--}$ do not exhibit photon echo. This can now be easily understood in terms of lensing. For these two susceptibilities, it is impossible to arrange the spin operators in such a way that all the created fractional excitations get annihilated at a later time. Our calculations indeed confirm this.

Our discussion so far is based on the ideal Luttinger spin liquid at the RG fixed point. In general, there are RG-irrelevant perturbations to the fixed point Hamiltonian Eq.~\eqref{eq:boson_hamil}, but they only give sub-leading corrections to most of the physical quantities at low temperatures. This also makes these RG-irrelevant corrections difficult to detect in experiments. Nevertheless, in the 2DCS in Luttinger spin liquid, the RG-irrelevant perturbations can produce pronounced effects because the lensing of fractional excitations relies on two features of the fixed point Hamiltonian: First, the phonon modes are the exact eigenstates of the Hamiltonian. Second, the phonon dispersion relation is exactly linear. These features ensure that a wave packet of fractional excitation can propagate indefinitely through the system without dissipation or dispersion. RG-irrelevant corrections to the fixed point Hamiltonian would prevent the indefinite propagation of the wave packet, and, therefore, could be sensitively detected by the suppression of lensing. 

In the XXZ spin chain, the RG-irrelevant perturbations include the umklapp terms such as $\cos(4\phi)$~\cite{Affleck1988,Giamarchi2003}, which result in the damping of phonon modes at finite temperature. Consequently, the wave packets of fractional excitations acquire finite life time. The lensing phenomenon shown in Fig.~\ref{fig:lensing} is suppressed when $\tau$ or $t_\mathrm{w}$ exceeds the life time of these excitations. This, in turn, will be manifest as the decay of the photon echo signal as $\tau$ or $t_\mathrm{w}$ increases, in close analogy with the dissipation-induced photon echo decay in few-body systems~\cite{Mukamel1995,Hamm2011}. Straightforward dimensional analysis suggests the decay rate vanishes as $T^{1+\eta}$ as $T\to0$ where $\eta$ is the dimension of the umklapp term.

On the other hand, there also exist a different family of RG-irrelevant perturbations, such as $(\nabla^2\phi)^2$, which produce a small curvature in the phonon dispersion relation while keeping them as the exact eigenstates of the Hamiltonian. As a result, the wave packet disperses as it propagates, but there is no true dissipation. The lensing picture suggests that the dispersion of wave packet would also result in the decay of the photon echo. For instance, the wave packet of the left-moving spinon created by $S^+(-3)$ would be quite dispersed when $\tau$ is large. It could not completely annihilate with the left-moving antispinon created by $S^{-}(-2)$ as the latter's wave packet is still sharp. This would spoil lensing. Thus, the lensing picture suggests a dispersion-induced decay mechanism for the photon echo in the Luttinger spin liquid. In the next section, we examine this mechanism in more detail. 

\section{Dispersion-induced decay \label{sec:dispersion}}

\begin{figure}
\centering
\includegraphics[width=\columnwidth]{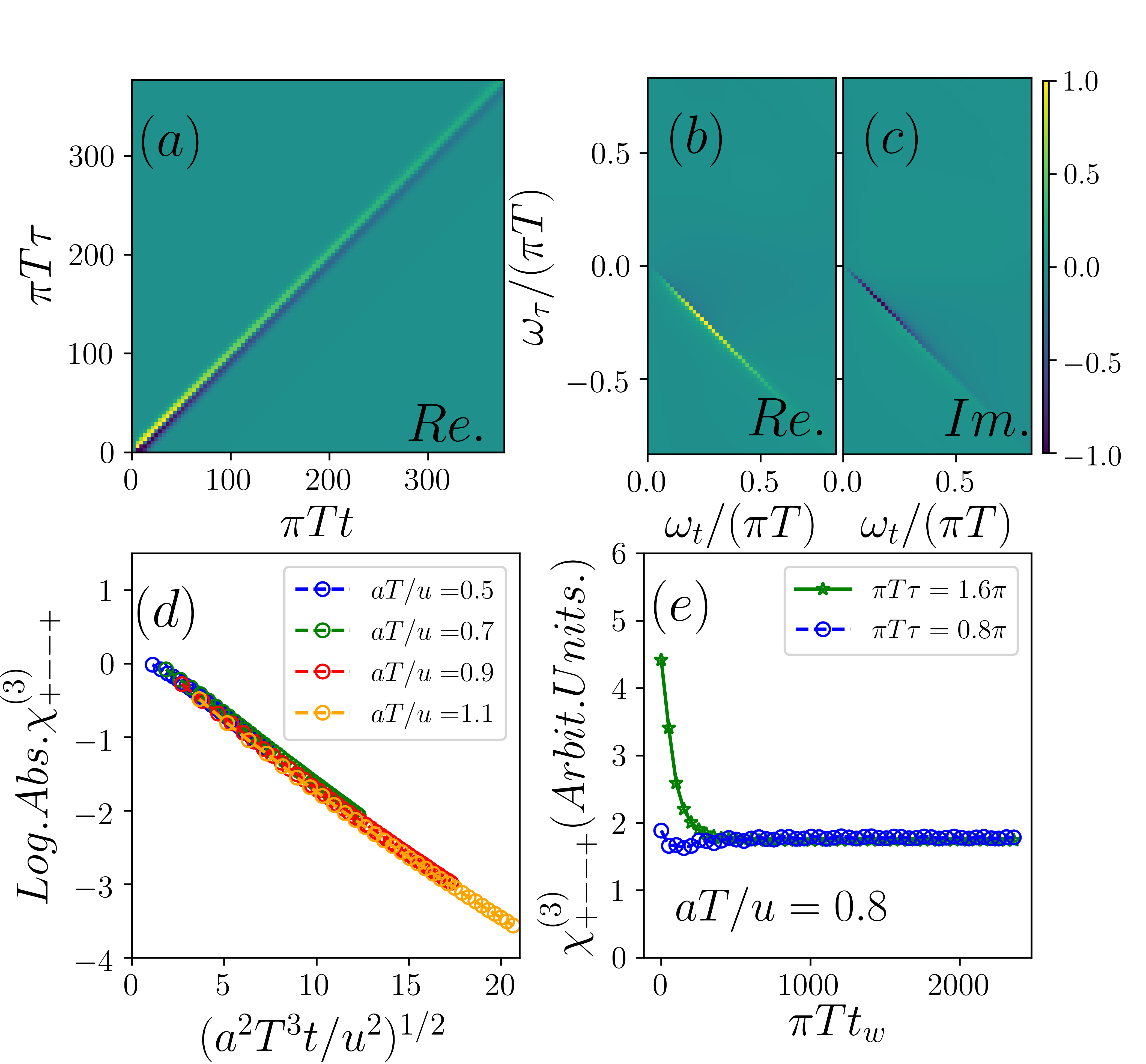}
\caption{(a) The nonlinear magnetic susceptibility $\chi^{(3)}_{+--+}$ as a function of $t$ and $\tau$, computed from the harmonic chain model Eq.~\eqref{eq:lattice_boson_hamil}. The waiting time $t_\mathrm{w} = 4a/u$. The temperature $T = 1.2u/a$. The data are scaled such that the maximum is 1. (b)(c) The real and imaginary parts of the two-dimensional spectrum obtained by Fourier-transforming the data in panel (a). Only the first and the fourth quadrants are shown. The other two quadrants are obtained by complex conjugation. (d) $\chi^{(3)}_{+--+}$, measured at $t=\tau$, as a function of $(a^2Tt/u^2)^{1/2}$ for various temperatures. $t_\mathrm{w} = 4a/u$. (e) $\chi^{(3)}_{+--+}$, measured at $t=\tau$, as a function of waiting time $t_\mathrm{w}$ for two representative values of $\tau$. The temperature $T=0.8u/a$.}
\label{fig:lattice_boson}
\end{figure}

The spinon lensing picture suggests that the dispersion of the wave packets could also lead to the decay of photon echo. In this section, we put this idea to test by a numerical experiment. We consider the harmonic chain, which is a discretization of the fixed point Hamiltonian:
\begin{align}
H = \frac{u}{2\pi a}\sum_{n} \frac{(\phi_{n+\frac{1}{2}}-\phi_{n-\frac{1}{2}})^2}{K}+ K (\theta_{n+1}-\theta_{n})^2 .
\label{eq:lattice_boson_hamil}
\end{align}
Here, $a$ is the lattice constant. $\theta_n$ resides on the lattice site labeled by $n$, whereas $\phi_{n+1/2}$ resides on the midpoint of the lattice link connecting the site $n$ and $n+1$. Their commutation relation is given by: $[\phi_{n+1/2},\theta_{n'}] = -i\pi\,\Theta(n-n'+1/2)$. It is sufficient for our purpose to consider the ferromagnetic chain where $S^\pm_n = \exp(\mp i\theta_n)$. We set the Luttinger parameter $K=1$.

The phonon modes are exact eigenstates of Eq.~\eqref{eq:lattice_boson_hamil} owing to its quadratic form. The phonon dispersion relation is now nonlinear: $\omega_q = 2u/a\,|\sin(qa/2)|$. Therefore, Eq.~\eqref{eq:lattice_boson_hamil} represents an idealized model system to study the dispersion-induced decay without the complications of dissipation effects. By contrast, in a microscopic spin lattice model, both effects are present and difficult to disentangle.

Fig.~\ref{fig:lattice_boson}(a) shows $\chi^{(3)}_{+--+}$ as a function of $t$ and $\tau$ at temperature $T=1.2u/a$. The waiting time $t_\mathrm{w}=4a/u$. Note the data are strictly real as the case in continuum (Fig.~\ref{fig:ana_fm}). We find a clear signature of photon echo running along the diagonal of the $(t,\tau)$ plane. However, the echo signal decays slowly along the diagonal direction owing to the dispersion effect. In the frequency domain, this decay manifests itself as a slight broadening of the photon echo peaks along the anti-diagonal direction of the frequency plane (Fig.~\ref{fig:lattice_boson}(b)\&(c)).

We now investigate the temperature dependence of the dispersion-induced decay. Expanding the phonon dispersion near $q=0$, we obtain $\omega_q = u(|q|-a^2 |q|^3/24+\cdots)$. Therefore, the width of the wave packet grows as $(a^2 ut)^{1/3}$ as it propagates. At temperature $T$, this defines a dispersion time scale $\tau_\mathrm{disp}$:
\begin{align}
(a^2 u\tau_\mathrm{disp} )^{1/3} \sim \xi \sim \frac{u}{T};\Rightarrow
\tau_\mathrm{disp}\sim \frac{u^2}{a^2 T^3}.
\end{align}
Here, $\xi$ is the spin correlation length. Beyond this time scale, the wave packet is essentially indistinguishable from thermal fluctuations. Thus, we hypothesize that the decay of the echo signal is controlled by $\tau_\mathrm{disp}$.

Our data support this hypothesis. Fig.~\ref{fig:lattice_boson}(c) shows the semi-log plot of $\chi^{(3)}_{+--+}$ at $t=\tau$ as a function of $(\tau/\tau_\mathrm{disp})^{1/2}$. The data for different $T$ approximately collapse to a straight line. This suggests the echo signal decays as a stretched exponential, $\exp(-C (\tau/\tau_\mathrm{disp})^\nu)$, where the exponent $\nu = 1/2$ and $C$ is a numeric constant. The decay rate $1/\tau_\mathrm{disp}\sim T^3$, consistent with the dimension  $\eta=2$ of the higher derivative term $(\nabla^2\phi)^2$. The origin of the stretching exponent $\nu = 1/2$ is unclear at the moment but likely associated with the asymptotic behavior of Airy functions.

Having investigated the decay of the echo as a function of $\tau$, we turn to its $t_\mathrm{w}$ dependence. Fig.~\ref{fig:lattice_boson}(e) shows $\chi^{(3)}_{+--+}$ at $t=\tau$ as a function of $t_\mathrm{w}$ for two representative values of $\tau$. The echo signal initially decrease as $t_\mathrm{w}$ increases, reflecting the effect of dispersion, but eventually saturate to a finite value This is also consistent with the lensing picture. A straightforward calculation shows that the spacetime-dependent susceptibility $\widetilde{\chi}_{+--+}$ saturates to finite value as $t_\mathrm{w}\to\infty$ at the lensing configuration Eq.~\eqref{eq:lensing_config}. In a few-body system, the decay of photon echo as a function of $t_\mathrm{w}$ reflects the thermal relaxation of the optical population. Here, as the phonon modes are exact eigenstates of the Hamiltonian Eq.~\eqref{eq:lattice_boson_hamil}, the phonon population cannot relax. We therefore heuristically attribute the saturation to the absence of thermal relaxation in Eq.~\eqref{eq:lattice_boson_hamil}.

To recapitulate, by a numerical experiment, we have shown that the photon echo signal decays as a function of $\tau$ in the presence of wave packet dispersion. The decay is controlled by a dispersion time scale $\tau_\mathrm{disp}$. Meanwhile, the echo signal saturates as $t_\mathrm{w}\to\infty$, which we attribute to the lack of thermalization in the model system.

\section{Discussion \label{sec:discussion}}

In this work, we have shown that the nonlinear magnetic susceptibility $\chi^{(3)}_{+--+}$ and its complex conjugate $\chi^{(3)}_{-++-}$ of the Luttinger spin liquid exhibit photon echo that resembles the perfectly rephasing photon echo in a few-body problem. However, the rephasing picture does not directly apply to the present system in that its energy spectrum is continuous. Instead, the echo signal arises from the lensing of the fractional excitations, and its decay as a function of the pulse delays is a sensitive diagnostic for the dissipation and/or dispersion.

The photon echo can be directly measured by THz 2DCS on spin chain materials that host Luttinger spin liquids. For example, Cs$_2$CoCl$_4$ is thought to be a realization of the easy-plane antiferromagnetic ($J_\perp>J_z>0$) $S=1/2$ XXZ chain~\cite{Berunig2013}. Although our analysis is carried out in the circular polarization basis, one could simplify the experimental set up by measuring the $\chi^{(3)}$ response with linear polarization (e.g. $\chi^{(3)}_{xxxx}$) since the linear polarization basis and the circular polarization basis are related by a linear transformation.

We find the lensing of fractional excitations to be a convenient picture for understanding the photon echo in the Luttinger spin liquid. A crucial feature of the lensing is the refocusing of the wave packets world lines, reminiscent of the refocusing of quantum phase accumulation in the NMR spin echo or photon echo in few-body systems. However, the lensing is unique to many-body system in that it entails the propagation of wave packets. It could be viewed as a conceptual extension of the more familiar interference picture~\cite{Mukamel1995,Hamm2011} commonly used in the study of photon echo in few-body systems from the time domain to the spacetime domain. 

An earlier work on the photon echo of quantum Ising chain uses the time-domain interference picture~\cite{Wan2019}, which is made possible by a mathematical mapping that relates the nonlinear response of a many-body system to that of an ensemble of independent few-body systems. Although it might be possible to adapt this methodology for the Luttinger spin liquid, we think it will be much less illuminating given the complex structure of the optical matrix elements.

An interesting consequence of the lensing picture is that the dispersion of the wave packet alone could lead to the decay of photon echo signal. This decay mechanism finds no immediate analogue in few-body systems. In the presence of dispersion but no dissipation, the echo signal decays as $\tau$ increases and eventually disappears when $\tau$ far exceeds the dispersion time scale. However, the echo signal does not disappear when the waiting time $t_\mathrm{w}\to\infty$. This points toward different uses of the pulse delays $\tau$ and $t_\mathrm{w}$ --- the former could be used as a dial to monitor both dispersion and dissipation effects, whereas the latter is sensitive to dissipation alone.

Our work opens a few directions worth further investigation. First of all, our discussion on the umklapp terms, or more broadly dissipation effects, has been qualitative. It will be useful to examine their impact on the photon echo quantitatively by employing the quantum kinetic theory~\cite{Tavora2013,Tavora2014}. Secondly, it would be interesting to compare our predictions with lattice model calculations. Our preliminary results on the XY spin chain~\cite{Lieb1961,McCoy1971,Derzhko1998,Derzhko2000,Maeda2003} show good agreement with the bosonization predictions and will be published elsewhere. Thirdly, our analysis focuses on the gapless phase of the XXZ type spin chain. It would be interesting to explore the nonlinear photon echo in the gapped phase by using a semiclassical treatment~\cite{Sachdev1997,Damle1998,Damle2005}, or in the vicinity of the Heisenberg point where the proximate $SU(2)$ symmetry might gives rise to new universal features~\cite{Oshikawa1999,Oshikawa2002}. We also note an interesting recent work where the generalized hydrodynamics is employed to compute some nonlinear responses of integrable systems including the XXZ spin chain~\cite{Fava2021}. Finally, even though we exclusively consider the spatially uniform (momentum $q=0$) magnetic response in this work, our calculations can be generalized to finite $q$. It has been shown that, in a magnetized antiferromagnetic Heisenberg chain, the magnetic resonance mode acquires a nonlinear dispersion at $q\neq0$ due to spinon interactions~\cite{Keselman2020,Kohno2009}. In light of these results, we expect that the $q\neq0$ nonlinear magnetic resonance may exhibit rich physics. Although optical measurements usually probes the $q=0$ response, the presence of  Dzyaloshinskii-Moriya (DM) interaction in a spin chain may allow for optical measurement of the $q \neq 0$ responses. The DM interaction may be eliminated by a gauge transformation at the expense of a momentum boost~\cite{OshikawaAffleck1997}. Hence, the $q=0$ response of the spin chain with a uniform DM interaction is equivalent to the finite $q$ response of the corresponding spin chain without the DM interaction~\cite{Bocquet2001,Gangadharaiah2008}.

Looking beyond the Luttinger spin liquids, we think our method and the physical picture could be applicable to other one-dimensional critical systems and potentially higher dimensional systems. Perhaps the most experimentally relevant are gapless systems with charged excitations, where the charge degrees of freedom could directly couple to the electric component of the THz pulse and therefore produce stronger nonlinear response. In short, we believe future investigation on the nonlinear response of many-body systems will uncover far richer dynamical phenomena and offer deeper insight  into these systems.

\begin{acknowledgments}
YW and MO thank the hospitality of the Kavli Institute for Theoretical Physics, University of California, Santa Barbara, where a part of this work was performed. 
MO also thanks Shunsuke Furukawa, Shunsuke C. Furuya, and Masahiro Sato for useful comments.
This work was supported in part by the National Natural Science Foundation of China (Grant No.~11974396), the Strategic Priority Research Program of the Chinese Academy of Sciences (Grant No.~XDB33020300), Japan Society for the Promotion of Science (Grant Nos.~JP18H03686 and JP19H01808), Japan Science and Technology Agency CREST (Grant No.~JPMJCR19T2), and the National Science Foundation (Grant No.~NSF PHY-1748958).  A part of the numerical calculations were carried out on Tianhe-1A  at the National Supercomputing Center in Tianjin, China. 
\end{acknowledgments}

\appendix

\section{Commutation relation prescriptions\label{app:prescription}}

In this appendix, we discuss the different choices for the commutation relation between $\phi$ and $\theta$ fields, and the subtle issues~\cite{DelftSchoeller1998} associated with these choices. Although elementary, these issues are pertinent to the calculation of higher order response functions. 

In bosonization, $\phi$ and $\nabla\theta$ form a pair of canonical conjugate variables:
\begin{align}
[\phi(x),\nabla\theta(y)] = i\pi\delta(x-y).
\end{align}
The above does not uniquely fix the commutator between $\phi$ and $\theta$. Several popular choices exist in literature. In this work, we use:
\begin{align}
[\phi(x),\theta(y)] = -i\pi\Theta(x-y).
\label{eq:app:heaviside}
\end{align}
We term this choice the ``Heaviside prescription" for latter convenience. A closely related variant is $[\phi(x),\theta(y)] = i\pi\Theta(y-x)$~\cite{Affleck1988}. Since the two variants are rather similar, we shall focus the prescription Eq.~\eqref{eq:app:heaviside}. Another popular choice is~\cite{Giamarchi2003}:
\begin{align}
[\phi(x),\theta(y)] = -i\frac{\pi}{2}\mathrm{Sgn}(x-y).
\end{align}
We term this choice the ``signum prescription".

In what follows, we show that the different prescriptions result in different bosonization dictionaries. We first consider the fermion field operator. For non-interacting fermions (Luttinger parameter $K=1$), the left and right chiral boson fields are dynamically coupled:
\begin{align}
\phi_L = \theta+\phi,\quad \phi_R = \theta-\phi.
\end{align}
The commutation relation of $\phi_L$, and likewise $\phi_R$, are identical for both prescriptions. However, the commutator between $\phi_L$ and $\phi_R$ depends on the prescription:
\begin{align}
[\phi_L(x),\phi_R(y)] = \left\{\begin{array}{cc}
-i\pi & (\mathrm{H.}) \\
0 & (\mathrm{S.})
\end{array}\right. .
\label{eq:app:LR_commutator}
\end{align}
In other words, boson fields with different chiralities do not commute (commute) in the Heaviside (signum) prescription. As a result, in the Heaviside prescription, the left and right chiral fermions:
\begin{subequations}
\begin{align}
\psi_L \sim e^{i\phi_L};\quad \psi_R \sim e^{i\phi_R},\quad(\mathrm{H.})
\end{align}
anti-commute thanks to the commutator between $\phi_L$ and $\phi_R$. By contrast, in the signum prescription, Klein factors are needed to ensure the anti-commutation relation:
\begin{align}
\psi_L \sim \eta_L e^{i\phi_L};\quad \psi_R \sim \eta_R e^{i\phi_R},\quad(\mathrm{S.}),
\end{align}
\end{subequations}
where $\eta_{L,R}$ are the Klein factors obeying the Clifford algebra: $\eta^2_L = \eta^2_R = 1$, and $\eta_L\eta_R = -\eta_R\eta_L$.

It is important to bear in mind that choosing different prescriptions do not change the dynamics of the boson fields. Furthermore, including fermion interactions ($K\neq1$) does not spoil the commutator between the left and right chiral boson fields. In this case, the two dynamically decoupled fields read:
\begin{align}
\phi_L = \sqrt{K}\theta+\frac{\phi}{\sqrt{K}},\quad \phi_R = \sqrt{K}\theta-\frac{\phi}{\sqrt{K}},
\end{align}
and we find Eq.~\eqref{eq:app:LR_commutator} holds regardless the value of $K$.

In the next, we consider the Jordan-Wigner string operator, which plays a crucial role in the bosonization of the XXZ spin chain:
\begin{align}
\mathcal{S}_j = \cos(\pi\sum_{n<j}c^\dagger_n c^{\phantom\dagger}_n),
\end{align}
where $c_n$ is the fermion annihilation operator on lattice site $n$. In particular, 
\begin{align}
c_i \mathcal{S}_j = (-)^{\Theta(j-i-1/2)} \mathcal{S}_j c_i.
\label{eq:app:jw_albegra}
\end{align}
We now bosonize the string operator following the standard procedure~\cite{Affleck1988,Giamarchi2003}: 
\begin{align}
\mathcal{S}_j &\approx \cos(k_Fx-\int^x_{-\infty}\nabla\phi(y)dy)
\nonumber\\
& = \cos(k_Fx-\phi(x)+\phi(-\infty)).
\end{align}
In the Heaviside prescription, the boundary term $\phi(-\infty)$ commutes with $\theta(x)$. We thus may regard it as a number and omit it:
\begin{subequations}
\begin{align}
\mathcal{S}_j = \cos(k_Fx-\phi(x)).\quad (\mathrm{H.}).
\end{align}
By contrast, in the signum prescription, we must keep the boundary term as $[\phi(-\infty),\theta(x)]\neq 0$:
\begin{align}
\mathcal{S}_j = \cos(k_Fx-\phi(x)+\phi(-\infty)).\quad (\mathrm{S.}).
\end{align}
\end{subequations}
In the signum prescription, the boundary term $\phi(-\infty)$ is needed to reproduce the commutation relation between the fermion field operator and the string operator (Eq.~\ref{eq:app:jw_albegra}). Had we dropped the boundary term, we would have found:
\begin{align}
c_i \mathcal{S}_j \overset{!}{=}e^{-i\pi/2\mathrm{Sgn}(x-y)}\mathcal{S}_jc_i,
\end{align}
which disagrees with Eq.~\ref{eq:app:jw_albegra}.

Finally, we are ready to bosonize the spin operators of the XXZ chain using the bosonization expressions of the fermion operator and the string operator. We assume the chain is antiferromagnetic ($J_\perp<0$) without loss of generality. Following Refs.~\onlinecite{Affleck1988,Giamarchi2003} but keeping a close eye on the Klein factors and the boundary term, we find the staggered components of the spin operators are given by:
\begin{align}
N^z_j \sim \left\{\begin{array}{cc}
\cos(2\phi) & (\mathrm{H.}) \\
 \eta_L\eta_R e^{-2i\phi}+\mathrm{H.C.} & (\mathrm{S.})
\end{array}\right. ,
\label{eq:Nz_boson}
\end{align}
\begin{align}
N^-_j \sim \left\{\begin{array}{cc}
e^{i\theta} & (\mathrm{H.}) \\
\eta_L e^{i(\theta+\phi(-\infty))}+\eta_R e^{i(\theta-\phi(-\infty))} & (\mathrm{S.})
\end{array}\right. .
\label{eq:N-_boson}
\end{align}
We stress that the Klein factors and the boundary term are essential in reproducing the correct commutation relations. For instance, had we dropped the Klein factors and the boundary term, we would have obtained:
\begin{align}
N^z(x) N^-(y) \overset{!}{=} -N^-(y) N^z(x),
\label{eq:app:anti-commute}
\end{align}
using the signum prescription. This is incorrect because these two operators must commute.

For the uniform components, we have:
\begin{align}
M^z_j \sim -\frac{1}{\pi}\nabla\phi,
\end{align}
which is independent of the prescription, and
\begin{align}
M^-_j \sim \left\{\begin{array}{cc}
e^{i\theta}\cos(2\phi) & (\mathrm{H.}) \\
\eta_L e^{i(\theta+2\phi-\phi(-\infty))}+\eta_R e^{i(\theta-2\phi+\phi(-\infty))} & (\mathrm{S.})
\end{array}\right. .
\label{eq:M-_boson}
\end{align}

We see that the bosonization dictionary takes a much simpler form in the Heaviside prescription. In fact, the Heaviside prescription has been used in the literature~\cite{HikiharaFurusaki2004,TeoKane2014}
when the precise bosonization formulae are needed. On the other hand, the full bosonization formulae of the spin operators in the $S=1/2$ chain in the signum prescription given in Eqs.~\eqref{eq:Nz_boson},~\eqref{eq:N-_boson}, and~\eqref{eq:M-_boson} are new to the best of our knowledge. For the signum prescription, one might hope that we could omit the Klein factors and the boundary term in calculations and still get correct results. This is indeed the case when calculating two-point functions, which is a fortunate coincidence. In fact, dropping the Klein factors and the boundary term from the bosonization dictionary can lead to incorrect results in the signum prescription when calculating higher order response functions.

To illustrate this point, consider the following four-point response function:
\begin{multline}
G^R(1,2,3) = i^3\Theta(t_1)\Theta(t_2-t_1)\Theta(t_3-t_2)
\\
 \times \langle[[ [N^x(0),N^z(-1)],N^x(-2)],N^z(-3)]\rangle.
\end{multline}
Had we dropped the Klein factors and the boundary term in the Signum prescription, we would have found $N^x$ and $N^z$ anti-commute when they are spacelike separated (Eq.~\eqref{eq:app:anti-commute}). As a result, $G^R\neq 0$ even when the point of detection is outside the light cone of the perturbation, thereby violating the causality principle. 

To recapitulate, different choices for the commutation relation of $\theta$ and $\phi$ lead to different bosonization dictionaries. The bosonization dictionary in the signum prescription includes Klein factors and the boundary term, which cannot be omitted in calculating the high-order response functions.

\section{Causality of the response function \label{app:causality}}
\begin{figure}
\includegraphics[width = \columnwidth]{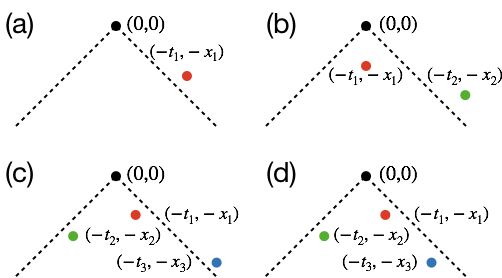}
\caption{(a)(b)(c) illustrates three possible cases in which the spacetime points $(-t_1,-x_1)$ (red dot), $(-t_2,-x_2)$ (green dot), $(-t_3,-x_3)$ (blue dot) lie outside of the past light cone (dashed line) of the origin (black dot). In any of the three cases, the response function $G^R$ (Eq.~\eqref{eq:gr_def}) vanishes. (d) Causality requires that all three points to lie inside the past light cone of the origin to induce a non-zero response.}
\label{fig:causality}
\end{figure}

In this appendix, we prove the causality of the response function $G^R$ (Eq.~\eqref{eq:gr_def}).

To set the stage, we recall that causality in relativistic quantum field theory requires that two observables commute if their separation is spacelike:
\begin{align}
[O_1(t_1,x_1),O_2(t_2,x_2)] = 0,\quad (u^2t_{12}^2 - x^2_{12}<0)
\label{eq:app:causal_condition}
\end{align}
where $O_{1,2}$ are arbitrary observables.

We first prove the following general statement: $G^R$ is causal provided that all the operators that appear in Eq.~\eqref{eq:gr_def} fulfil the condition Eq.~\eqref{eq:app:causal_condition}. Specifically, we need to show $G^R=0$ if any of the three spacetime points $(-t_1,-x_1)$, $(-t_2,-x_2)$, $(-t_3,-x_3)$ are outside the light cone of $(0,0)$. We prove this statement by exhaustion: 
\begin{enumerate}[label=(\roman*)]
\item If $(-t_1,-x_1)$ is outside the light cone of $(0,0)$ (Fig.~\ref{fig:causality}(a)), then Eq.~\eqref{eq:app:causal_condition} implies $[V_0,V_1] = 0$, and consequently $G^R=0$. 

\item Now suppose $(-t_1,-x_1)$ is inside the light cone of $(0,0)$. If $(-t_2,-x_2)$ is outside the light cone of $(0,0)$, it must also be outside the light cone of $(-t_1,-x_1)$  (Fig.~\ref{fig:causality}(b)). Thus, $[V_0,V_2] = [V_1,V_2] = 0$ by virtue of Eq.~\eqref{eq:app:causal_condition}. It then follows that $[[V_0,V_1],V_2] = 0$, and consequently $G^R=0$.

\item Now suppose both $(-t_1,-x_1)$ and $(-t_2,-x_2)$ are inside the light cone of $(0,0)$. If $(-t_3,-x_3)$ is outside the light cone of $(0,0)$, it must also be outside the light cone of $(-t_1,-x_1)$ and $(-t_2,-x_2)$ (Fig.~\ref{fig:causality}(c)). Thus, $[V_0,V_3] = [V_1,V_3] = [V_2,V_3]= 0$ thanks to Eq.~\eqref{eq:app:causal_condition}. It follows that $[[[V_0,V_1],V_2],V_3] = 0$, and thus $G^R=0$.
\end{enumerate}
This completes our proof.

In the next, we turn to the Tomonaga-Luttinger liquid theory and show that the local vertex operators:
\begin{align}
V = e^{im\theta+2in\phi},\quad (m,n\in\mathbb{Z})
\end{align}
indeed fulfill the condition Eq.~\eqref{eq:app:causal_condition}. To this end, we consider two vertex operators $V_1$ and $V_2$. Suppose they are spacelike separated. We choose a reference frame in which they are synchronous and compute their equal time commutator:
\begin{align}
V_1V_2 &= e^{i (m_1\theta(x_1)+2n_1\phi_(x_1))}e^{i(m_2\theta(x_2)+2n_2\phi_(x_2))}
\nonumber\\
&= e^{2\pi i (n_1m_2\Theta(x_1-x_2)-m_1n_2\Theta(x_2-x_1) )}V_2V_1
 \nonumber\\
&= V_2 V_1. 
\end{align}
The second equality follows from the Baker-Campbell-Hausdorff formula; the third equality follows from the fact $m_{1,2},n_{1,2}\in\mathbb{Z}$. The above immediately implies:
\begin{align}
[V_1(x_1),V_2(x_2)] = 0.
\end{align}
Thus, we have verified that the local vertex operators fulfill the causality condition Eq.~\eqref{eq:app:causal_condition}. 

Combining the above results, we conclude that the response function $G^R$ (Eq.~\eqref{eq:gr_def}) is causal. We note that one could also verify the causality of $G^R$ by using its explicit expression Eq.~\eqref{eq:gr_result}.

\section{Antiferromagnetic Heisenberg chain \label{app:heisenberg}}

In this appendix, we compute $\chi^{(3)}_{+--+}$ of the antiferromagnetic Heisenberg chain ($J_z = J_\perp>0$ in Eq.~\eqref{eq:xxz_hamil}) by using the equation of motion of the lattice model~\cite{Oshikawa1999,Oshikawa2002}. 

The equation of motion for the total magnetization $M^\pm = M^x \pm i M^y$ reads:
\begin{align}
i \frac{\partial M^\pm}{\partial t} = [M^\pm,H] = [M^\pm,-BM^z] = \pm BM^\pm.
\end{align}
The second equality follows from the fact that the Heisenberg interaction, being $SU(2)$ symmetric, commutes with $M^\pm$. Solving the above, we find:
\begin{align}
M^\pm(t) = e^{\mp i Bt}M^\pm.
\end{align}
Substituting the above into the Kubo formula for $\chi^{(3)}_{+--+}$, and using the spin commutation relation, we obtain:
\begin{align}
\chi^{(3)}_{+--+} = -4i m \Theta(t)\Theta(t_\mathrm{w})\Theta(\tau)e^{-iB(t-\tau)},
\end{align}
where $m$ is the magnetization density. Therefore, for the Heisenberg chain, the nonlinear magnetic susceptibility shows oscillation with constant magnitude, which reflects the Larmor precession of the total magnetization vector. We stress that this simplicity is unique to the spatially uniform (momentum $q=0$) response function. By contrast, the $q\neq0$ magnetic response functions may exhibit rich physics that is beyond the Larmor precession. For instance, the dispersion relation of the magnetic resonance mode, emanating from the Larmor frequency $\omega = B$ at $q=0$, acquires a curvature thanks to the spinon interactions~\cite{Keselman2020,Kohno2009}.  These effects are not visible in the $q=0$ response function.


\bibliography{xxz_nonlinear.bib}
\end{document}